\magnification=1200
\hoffset=.0cm
\voffset=.0cm
\baselineskip=.55cm plus .55mm minus .55mm

%
%
%
\input amssym.def
\input amssym.tex
%
%
%
%
%
%
%
%
%


\font\grassettogreco=cmmib10
\font\scriptgrassettogreco=cmmib7
\font\scriptscriptgrassettogreco=cmmib10 at 5 truept
\textfont13=\grassettogreco
\scriptfont13=\scriptgrassettogreco
\scriptscriptfont13=\scriptscriptgrassettogreco


\font\sansserif=cmss10
\font\scriptsansserif=cmss10 at 7 truept
\font\scriptscriptsansserif=cmss10 at 5 truept
\textfont14=\sansserif
\scriptfont14=\scriptsansserif
\scriptscriptfont14=\scriptscriptsansserif
\def\sans{\fam=14}


\font\capital=rsfs10
\font\scriptcapital=rsfs10 at 7 truept
\font\scriptscriptcapital=rsfs10 at 5 truept
\textfont15=\capital
\scriptfont15=\scriptcapital
\scriptscriptfont15=\scriptscriptcapital
\def\scri{\fam=15}


\font\euler=eusm10
\font\scripteuler=eusm7
\font\scriptscripteuler=eusm5 
\textfont12=\euler
\scriptfont12=\scripteuler
\scriptscriptfont12=\scriptscripteuler

\font\cps=cmcsc10
%

%
%
%
%
%

%
%
%
%
%
%
\def\ref#1{\lbrack#1\rbrack}
%
%
%
%
%
\def\dim{{\rm dim}\hskip 1pt}

\def\imu{\sqrt{-1}}

\def\ker{{\rm ker}\hskip 1pt}
\def\ran{{\rm ran}\hskip 1pt}

\def\tr{{\rm tr}\hskip 1pt}

\def\ad{{\rm ad}\hskip 1pt}
\def\Ad{{\rm Ad}\hskip 1pt}

\def\SO{{\rm SO}}
\def\SU{{\rm SU}}

\def\Gau{{\rm Gau}\hskip 1pt}

\def\Tor{{\rm Tor}\hskip 1pt}

\def\Conn{{\rm Conn}\hskip 1pt}

\def\hst1{\hskip 1pt}
\def\square{\,\vbox{\hrule \hbox{\vrule height 0.25 cm 
\hskip 0.25 cm \vrule height 0.25 cm}\hrule}\,}
\def\ul#1{\underline{#1}{}}
%
%
%
%
%

\def\defn#1{\vskip.3cm${\hbox{\cps Definition {#1}}}~~$}
\def\prop#1{\vskip.3cm${\hbox{\cps Proposition {#1}}}~~$}
\def\defprop#1{\vskip.2cm${\hbox{\cps Definition and Proposition {#1}}}~~$}
\def\theo#1{\vskip.3cm${\hbox{\cps Theorem {#1}}}~~$}

\def\rema#1{\vskip.3cm${\hbox{\cps Remark {#1}}\vphantom{f}}~~$}
\def\exa#1{\vskip.3cm${\hbox{\cps Example {#1}}\vphantom{f}}~~$}
\def\proof{\vskip.3cm${\hbox{\it Proof}}~~~$}
\def\titlebf#1{\vskip.5cm${\underline{\hbox{\bf #1}}}$\vskip.5cm}
\def\titlecps#1{\vskip.3cm${\underline{\hbox{\cps #1}}}$\vskip.3cm}


\hrule\vskip.5cm
\hbox to 16.5 truecm{June 2003  \hfil }
\hbox to 16.5 truecm{Version 2  \hfil hep-th/0306287}
\vskip.5cm\hrule
\vskip.9cm
\centerline{\bf GLOBAL ASPECTS OF ABELIAN AND CENTER PROJECTIONS}  
\centerline{\bf IN SU(2) GAUGE THEORY}
\vskip.4cm
\centerline{by}
\vskip.4cm
\centerline{\bf Roberto Zucchini}
\centerline{\it Dipartimento di Fisica, Universit\`a degli Studi di Bologna}
\centerline{\it V. Irnerio 46, I-40126 Bologna, Italy}
\centerline{\it and }
\centerline{\it INFN, sezione di Bologna}
\vskip.9cm
\hrule
\vskip.6cm
\centerline{\bf Abstract} 
We show that the global aspects of Abelian and center projection of a 
$\SU(2)$ gauge theory on an arbitrary manifold are naturally described 
in terms of smooth Deligne cohomology. This is achieved through the 
introduction of a novel type of differential topological structure,
called Cho structure. Half integral monopole charges appear naturally in 
this framework.
\vskip.4cm
\par\noindent


\vfill\eject

\vskip.6cm
\titlebf {Table of Contents}
\vskip.6cm
\hrule
\vskip.6cm

\item{1.} Introduction and conclusions

\item{2.} $\SU(2)$ principal bundles and associated adjoint vector bundles

\item{3.} Cho structures

\item{4.} The Deligne cohomology class $D_{\scri C}$

\item{5.} Fine and almost fine Cho structures and the Deligne class $D_{\scri C}$

\item{6.} Diagonalizability and almost fineness

\item{7.} Monopole and instanton charge, twist sectors and vortices

\item{A.} Smooth Deligne cohomology

\vfill\eject

\titlebf {1. Introduction and conclusions}

Since the early seventies, a considerable effort has been devoted to the 
theoretical understanding of confinement in non Abelian gauge theory.
The two most accredited theories are the dual superconductor model
\ref{1--6} and the vortex condensation model \ref{7--11}.

According to the dual superconductor model picture, confinement is due to
the condensation of chromomagnetic monopoles which forces the chromoelectric 
field into flux tubes, through a mechanism known as dual Meissner effect,
leading to a linear rising confining chromoelectric potential.

Abelian gauge fixing and Abelian projection were proposed by 't Hooft 
in 1980 as a means of explaining the emergence of monopoles in non Abelian 
gauge theory \ref{6}. In an Abelian gauge, such as the maximal and the 
Laplacian Abelian gauges, the gauge symmetry associated 
with the coset $G/H$ of the maximal Abelian subgroup $H$ of the gauge group 
$G$ is fixed. The resulting gauge fixed field theory is an Abelian gauge 
theory with gauge group $H$, in which the $H$ and $G/H$ components 
of the original gauge field 
behave effectively as Abelian gauge fields and matter fields, respectively. 
The Abelian projection consists in keeping the $H$ component of the gauge 
field and neglecting the $G/H$ one, separating out in this way the 
Abelian sector of the theory relevant for confinement.

The gauge transformation required to transform a given gauge field into 
one satisfying the appropriate Abelian gauge fixing condition is not smooth 
in general. (This is just another manifestation of the well known Gribov 
problem.) The transformed gauge field has therefore defects. 
In a $4$--dimensional space--time, the defects are located on 
closed loops. These are the chromomagnetic monopole world lines. 
The monopoles are remnants of the $G/H$ part of the gauge symmetry. 
Their condensation leads to confinement. 
Without the monopoles, Abelian projection would yield a rather trivial 
non confining Abelian gauge theory. 

In the vortex condensation model picture, confinement is induced by 
the filling of the vacuum by closed chromomagnetic center vortices. Their 
condensation leads to an area law for the Wilson loop and thus to confinement.

The emergence of vortices is explained by means of center gauge fixing 
and center projection \ref{12}. In a center gauge, such as the maximal and 
the Laplacian center gauges, the gauge symmetry 
associated with the coset $G/Z$ of the center $Z$ of the gauge group $G$
is fixed. Since $Z$ is typically a finite group, the resulting 
gauge fixed field theory is an exotic $Z$ gauge theory with no obvious 
continuum interpretation. The center projection consists in keeping the $Z$ 
degrees of freedom of the gauge field and neglecting the $G/Z$ ones.
The former are related to the global topological properties of the 
gauge field. 

As for Abelian gauge fixing, the gauge transformation required to transform 
a given gauge field into one satisfying the appropriate center gauge fixing 
condition is not smooth in general. The transformed gauge field has
therefore defects. In a $4$--dimensional space--time, the defects 
are located on closed surfaces. These are the
chromomagnetic center vortex world sheets. 
The center vortices are remnants of the $G/Z$ part of the gauge symmetry.
Their condensation is responsible for confinement. 

Since the dual superconductor model and the vortex condensation model 
describe the same physical phenomena, there should be a way of identifying
vortices in the dual superconductor picture and monopoles in the 
vortex condensation picture. This is indeed possible provided the gauge fixing 
is carried out appropriately as follows.

Laplacian center gauge fixing can be implemented in two steps \ref{13,14}.
In the first step, the gauge symmetry is partially fixed from $G$ to $H$.
In the second step, the gauge symmetry is further partially fixed from 
$H$ to $Z$. The first step is nothing but Abelian gauge fixing, which 
therefore becomes an integral part of center gauge fixing. Therefore, 
in this center gauge, the vortex defects of a gauge field 
include the monopole defects as a distinguished subset. Correspondingly, 
the vortex world sheets contain the associated monopole world 
lines. In this way, vortices appear as chromomagnetic flux tubes 
connecting monopoles \ref{15}.  

To take into account the topologically non trivial gauge fields 
corresponding to mono\-poles, it is necessary to allow for twisted boundary 
conditions \ref{16--18}.
In pure gauge theory, gauge transformation proceeds via the adjoint 
representation of $G$. Since the center $Z$ of $G$ lies in the kernel
of the latter, one may embed the original gauge theory in a 
broader one, by allowing the $G$ valued functions, in terms of which the boundary 
conditions obeyed by the gauge fields are expressed, to satisfy the appropriate 
$1$--cocycle conditions only up to a $Z$ valued twist.
There are several types of boundary conditions depending on the possible  
twist assignments. Correspondingly, the gauge fields fall in topological
classes or twist sectors. 
The gauge fields of the original untwisted theory form the trivial twist sector.
The gauge fields of the non trivial twist sectors are obtained from those of
the trivial twist sector by applying suitable multivalued gauge transformations
with $Z$ monodromy. The branching sheets of the latter are the vortex world sheets.  

\vfill\eject

\titlecps{Formulation of the problem}

One can view an Abelian gauge as map that assigns to each gauge field $A$ a non 
vanishing Higgs field $\phi(A)$ transforming in the adjoint representation of 
the gauge group $G$, i. e. satisfying 
$$
\phi(A^g)=\Ad g^{-1}\phi(A), 
\eqno(1.1)
$$
for any gauge transformation $g$, where $A^g$ is the gauge transform of $A$. By 
definition, a gauge transformation $g$ carries $A$ into the Abelian gauge if 
$\phi(A^g)$ is $\goth h$ valued \ref{6}. If $g$ has this property, then also $gh$ 
does, for any $H$ valued gauge transformation $h$. So, the Abelian gauge fixing 
leaves a residual unfixed $H$ gauge invariance. The defect manifold $N$ of the 
Abelian gauge is formed by those points $x$ of space time where $\phi(\bar A)(x)$ 
is invariant under the adjoint action of a subgroup $K$ of $G$ properly containing 
$H$, where $\bar A$ is any gauge field belonging to the Abelian gauge, 
whose choice is immaterial \ref{19,20}.
If $g$ is a gauge transformation carrying $A$
into the Abelian gauge, then $g$ must be singular at a Dirac manifold 
$D_g$ bounded by $N$ depending on $g$. Consequently, $A^g$, also, is. 
$N$ consists of monopole world lines and $D_g$ of the associated Dirac strings. 
The Abelian projection consists in replacing $A$ with either one of the following 
$$
A_{\rm sff}=\Pi_{\goth h}A^g,\qquad A_{\rm bff}=(\Pi_{\goth h}A^g)^{g^{-1}},
\eqno(1.2)
$$
where $\Pi_{\goth h}$ is a suitable projector of $\goth g$ onto $\goth h$. The 
two choices corresponds to the so called space and body fixed frame \ref{21}.
The body fixed frame form $A_{\rm bff}$, which we adopt in the following, 
has the advantage of being singular only at the defect manifold $N$.

The center gauge fixing works to some extent in similar fashion. It requires two 
linearly independent Higgs fields $\phi(A)$, $\phi'(A)$ satisfying (1.1) 
rather than a single one.
A gauge transformation $g$ carries $A$ into the center gauge if $\phi(A^g)$ is 
$\goth h$ valued, as before, and  $\phi'(A^g)$ is $\goth v$ valued, where $\goth v$ 
is a suitable proper subspace of $\goth g$ not contained $\goth h$ \ref{13,14}. 
If $g$ has this property, then also $gz$ does, for any element $z$ of $Z$.
So, the center gauge fixing leaves a residual unfixed $Z$ gauge invariance. 
The defect manifold $N^*$ of the center gauge is formed by those points $x$ of 
space time where $\phi(\bar A)(x)$, $\phi'(\bar A)(x)$ are simultaneously  
invariant under the adjoint action of a subgroup $K$ of $G$ properly containing 
$Z$, where $\bar A$ is any gauge field belonging to the center gauge \ref{19,20}.
If $g$ is a gauge transformation carrying $A$ into the center gauge, then $g$ must 
be singular at a Dirac manifold $D^*{}_g$ bounded by $N^*$ depending on $g$ and,
so, $A^g$ also is. 
$N^*$ consists of vortex world sheets and $D^*{}_g$ of the associated Dirac volumes.
The singularity of $g$ appears as a non trivial $Z$ monodromy around
$N^*$, here working as a branching manifold. The center projection 
replaces $A^g$ with $g^{-1}dg$. The latter has non trivial $Z$ valued holonomy
if $g$ has non trivial $Z$ monodromy \ref{22}

By allowing gauge transformations with non trivial $Z$ monodromy in center gauge 
fixing, one is effectively trading the original $G$ gauge theory by a $G/Z$ one 
possessing a richer wealth of topological classes. These correspond to the twist 
sectors discussed earlier. These sectors should somehow appear also in Abelian gauge 
fixing, since one expects the various gauge fixing procedures to lead eventually to 
physically equivalent descriptions of gauge theory low energy dynamics.

Below, we restrict to $\SU(2)$ gauge theory to allow for a particularly simple 
and direct treatment. Here, as is usual in the physical gauge theory
literature, we use the convenient isovector notation, in which the Lie algebra 
$\goth s\goth u(2)$ and its Cartan and Lie brackets are identified with 
the 3--dimensional Euclidean space $\Bbb E_3$ and its dot and cross product, 
respectively. Instead of the Higgs field $\ul\phi$, it is customary 
to employ the normalized Higgs field 
$$
\ul n=\ul\phi/|\ul\phi|, \qquad \ul n^2=1.
\eqno(1.3)
$$
The body fixed frame gauge field $\ul A_{\rm bff}$ is of the form 
$$
\ul A_{\rm bff}=a \ul n+d\ul n\times\ul n.
\eqno(1.4)
$$
The gauge field $\ul A_{\rm bff}$ was first written by Cho \ref{23,24}. 
It is characterized by the fact that $\ul n$ is covariantly constant 
with respect to it
$$
\ul D_{\rm bff}\ul n=d\ul n+\ul A_{\rm bff}\times\ul n=0.
\eqno(1.5)
$$
As a consequence, $\ul A_{\rm bff}$ is a reducible gauge field. 
In this case, the defect manifold $N$ of an Abelian gauge is simply 
the vanishing locus of the associated Higgs field $\ul\phi$. 
At the points of $N$, $\ul n$ is singular, as $\ul A_{\rm bff}$ is.
The defects represent monopoles. The monopole charge is given by
$$
m=-{1\over 4\pi}\int_\Sigma\ul F_{\rm bff}\cdot\ul n
={1\over 4\pi}\int_\Sigma\big(-da_{\rm bff}
+\hbox{$1\over 2$}\ul n\cdot d\ul n\times d\ul n\big),
\eqno(1.6)
$$
where $\ul F_{\rm bff}$ is the curvature of $\ul A_{\rm bff}$ and 
$\Sigma$ is a surface surrounding the monopoles.

The center of $\SU(2)$ is the group $\{\pm 1_2\}\cong \Bbb Z_2$.
For reasons discussed earlier, $\ul A_{\rm bff}$ is an 
$\SU(2)/\Bbb Z_2\cong\SO(3)$ gauge field rather than an $\SU(2)$ one.  
In general, $\ul A_{\rm bff}$ is not liftable to an $\SU(2)$ gauge field. 
The associated obstructions are related to center vortices. 

\titlecps{Plan of the paper}

The compatibility of Abelian and center projection with the global topology of 
the gauge background is not obvious a priori and requires a critical examination. 
In fact, the truncation of the gauge fields implicit in the projections,
though doable locally, is not manifestly so globally, when the background topology 
is non trivial. We ask then the following question. Is there a natural global
structure in which the local field theoretic data yielded by the projection can be 
fitted in? If so, which are its properties? 
The aim of this paper is to explore this matter.

The standard treatment of the defects associated with Abelian or center gauge 
fixing requires specific ad hoc choices of coordinates and trivializations 
making covariance obscure. Non trivial topology appears in the form of 
Dirac strings, sheets and the like emanating from the defects. 
We adopt an alternative approach avoiding this. 
It consists in defining all the fields locally and 
analyzing their gluing. Dirac strings, sheets etc. are then traded for cocycles 
specifying the gluing of the local fields. 
This is in the spirit of the seminal work of Wu and Yang \ref{25}.
Mathematically, it requires the apparatus of \v Cech cohomology \ref{26--28}. 
It is an alternative approach to the geometry and topology of these gauge 
models which relies on cohomology rather than homotopy. It has the 
advantage of being very general allowing for the analysis of Abelian and 
center projections for a gauge theory on an arbitrary space time manifold. 

To study the global features of the projections
in $\SU(2)$ gauge theory, we shall work out a local formulation of Cho's
gauge theoretic framework \`a la \v Cech \ref{23,24}. 
This will lead to a novel type
of differential topological structures called {\it Cho structures}.
In due course, we shall discover that the global aspects  
of a Cho structure $\scri C$ are encoded in a length $3$ degree $3$
Deligne cohomology class $D_{\scri C}$ \ref{29--32}, whose 
properties will be studied in detail. Next, we shall argue that 
a Cho structure $\scri C$ describes monopoles provided the associated 
Deligne class $D_{\scri C}$ vanishes and we shall show that, when 
this happens, the resulting monopole configurations are in one--to--one 
correspondence with certain differential topological structures
subordinated to $\scri C$, called {\it fine Cho structures}. We shall also 
prove that these latter are classified by the length $2$ degree 
$2$ Deligne cohomology (up to equivalence). 
Next, we shall show that {\it $D_{\scri C}$ does indeed always vanish}. 
This will be the main result of the paper. Finally, we shall also interpret 
center vortices as $\Bbb Z_2$ topological obstructions
to the lifting of the $\SO(3)$ bundles associated with the fine Cho
structures to $\SU(2)$ ones. {\it $\Bbb Z/2$ valued monopole charges 
and $\Bbb Z/4$ valued topological charges emerge naturally in this 
framework}.

\titlebf {2. $\SU(2)$ principal bundles and associated adjoint vector bundles}

In this paper, we consider exclusively principal $\SU(2)$ bundles
and the associated vector bundles. $\SU(2)$ is the simplest non Abelian 
group and one can conveniently take advantage of that.
We shall exploit throughout the following basic properties of $\SU(2)$.
The material expounded below is well known and is collected only 
to define the notation and the conventions used in the sequel of the paper. 
The isovector notation used is standard in the physical literature.

\titlecps{Generalities on $\SU(2)$}

The $3$--dimensional oriented Euclidean space $\Bbb E_3$ is a
Lie algebra. The Lie brackets and the Cartan form of $\Bbb E_3$
are given by $[\ul x,\ul y]=\ul x\times\ul y$ and 
$(\ul x,\ul y)=-2\ul x\cdot\ul y$, for $\ul x,\ul y\in\Bbb E_3$,
respectively. The map $\ul{\sans h}:\goth s\goth u(2)\rightarrow
\Bbb E_3$ defined by $\sigma(\ul {\sans h}(x))/2i=x$, for 
$x\in \goth s\goth u(2)$, where $\sigma$ denotes the 
standard Pauli matrices, is a Lie algebra isomorphism.
Hence, the Lie algebras $\goth s\goth u(2)$, $\Bbb E_3$
are isomorphic. The resulting identification is particularly convenient 
as the Lie brackets and the Cartan form of $\goth s\goth u(2)$
are represented by the ordinary cross and dot product of vectors
of $\Bbb E_3$, respectively. Below, we shall use extensively the convenient
shorthands $\ul x=\ul {\sans h}(x)$, for $x\in \goth s\goth u(2)$,
and $\star\,\ul x\ul y=-\ul{\sans h}(\ad xy)=-\ul x\times\ul y$, 
for $x, y\in\goth s\goth u(2)$. 

The center of $\SU(2)$ is $\Bbb Z_2$ realized as the sign group
$\{\pm 1\}$. The map $\ul{\sans H}:\SU(2)/\Bbb Z_2\rightarrow \SO(3)$ 
defined by $\sigma(\ul{\sans H}(R)\ul x)=\Ad R\sigma(\ul x)$, for
$R\in\SU(2)/\Bbb Z_2$, $\ul x\in \Bbb E_3$, is a Lie group isomorphism.
Hence, the groups  $\SU(2)/\Bbb Z_2$, $\SO(3)$ are isomorphic.
Correspondingly, we have an isomorphism 
$\ul{\dot{\sans H}}:\goth s\goth u(2)\rightarrow \goth s\goth o(3)$.
Hence, the Lie algebras $\goth s\goth u(2)$, $\goth s\goth o(3)$
are isomorphic. Below, we shall use the shorthand
$\ul R=\ul {\sans H}(R)$, for $R\in\SU(2)$.

\titlecps{Generalities on $\SU(2)$ principal bundles}

On a topologically non trivial manifold $M$, fields are sections of vector 
bundles. These, in turn, are associated with principal bundles and 
linear representations of their structure group.
\v Cech theory provides an advantageous description of these 
differential topological structures. It requires the 
introduction of a sufficiently fine open cover $\{O_\alpha\}$ of $M$
on whose sets the local representations of the fields are given.
\footnote{}{}\footnote{${}^1$}{Below, we denote by $\varphi_\alpha$
the local representation of a \v Cech $0$--cochain $\varphi$ on $O_\alpha$.
Similarly, we denote by $\varpi_{\alpha\beta}$ the local representation of a 
\v Cech $1$--cochain $\varpi$ on $O_\alpha\cap O_\beta\not=\emptyset$, etc.
For convenience, we shall suppress the indexes $\alpha$, $\beta$, ...
when confusion cannot arise.} See for instance \ref{33} for background 
material.

Let $P$ be an $\SU(2)$ principal bundle on a manifold $M$.
Then, $P$ is represented by some $\ul{\SU}(2)$ \v Cech $1$--cocycle 
$\{R_{\alpha\beta}\}$.

The adjoint bundle $\Ad P$ of $P$ is the vector bundle
$$
\Ad P=P\times_{\SU(2)}\goth s\goth u(2)
,\eqno(2.1)
$$
where $\SU(2)$ acts on $\goth s\goth u(2)$ by the adjoint representation. 
By the isomorphisms $\ul{\sans h}:\goth s\goth u(2)\rightarrow\Bbb E_3$
and ${\sans H}:\SU(2)/\Bbb Z_2\rightarrow\SO(3)$, $\Ad P$ is isomorphic to the 
oriented rank $3$ vector bundle $E$ 
$$
E=P\times_{\SU(2)}\Bbb E_3
,\eqno(2.2)
$$
where $\SU(2)$ acts on $\Bbb E_3$ via the representation induced by 
the natural map $\SU(2)\rightarrow\SU(2)/\Bbb Z_2$ and the
isomorphism $\ul{\sans H}$. $E$ is represented by the $\ul{\SO}(3)$ \v Cech 
$1$--cocycle $\{\ul R_{\alpha\beta}\}$ corresponding to the $\ul{\SU}(2)$ 
\v Cech $1$--cocycle $\{R_{\alpha\beta}\}$ under the isomorphism $\sans H$.
The vector bundle $E$ is not generic, since its first and second 
Stiefel--Whitney classes both vanish, $w_1(E)=0$, $w_2(E)=0$.
In the following, we find definitely more convenient to work with the vector 
bundle $E$ rather than the adjoint bundle $\Ad P$ of $P$, since this allows 
us to exploit familiar techniques of vector calculus. 

Let $A\in\Conn(M,P)$ be a connection of the principal bundle $P$. 
$A$ is represented by an $\Omega^1\otimes\goth s\goth u(2)$ \v Cech
$0$--cochain $\{A_\alpha\}$ satisfying
$$
A_\alpha=\Ad R_{\alpha\beta}A_\beta-d R_{\alpha\beta}R_{\alpha\beta}{}^{-1}
.\eqno(2.3)
$$
By the isomorphisms $\ul{\sans h}:\goth s\goth u(2)\rightarrow\Bbb E_3$
and $\ul{\sans H}:\SU(2)/\Bbb Z_2\rightarrow\SO(3)$,
$A$ induces a connection $\ul A\in \Conn(M,E)$ of the 
vector bundle $E$. $\ul A$ is represented by an $\Omega^1\otimes\Bbb E_3$ 
\v Cech $0$--cochain $\{\ul A_\alpha\}$ satisfying
$$
\ul A_\alpha=\ul R_{\alpha\beta}(\ul A_\beta-\ul\omega_{\alpha\beta})
,\eqno(2.4)
$$
where $\ul\omega_{\alpha\beta}$ is the $\Bbb E_3$ valued $1$--form defined by
$$
\star\,\ul\omega_{\alpha\beta}=-\ul R_{\alpha\beta}{}^{-1}d\ul R_{\alpha\beta}
.\eqno(2.5)
$$ 
This correspondence establishes a canonical isomorphism of the affine spaces 
$\Conn(M,P)$, $\Conn(M,E)$ of connections of $P$, $E$.

Let $s\in\Omega^p(M,\Ad P)$ be a $p$--form section of $\Ad P$.
Then, $s$ is represented by an $\Omega^p\otimes\goth s\goth u(2)$ \v Cech 
$0$--cochain $\{s_\alpha\}$ matching as 
$$
s_\alpha=\Ad R_{\alpha\beta}s_\beta
.\eqno(2.6)
$$
By the isomorphisms $\ul{\sans h}:\goth s\goth u(2)\rightarrow\Bbb E_3$
and $\ul{\sans H}:\SU(2)/\Bbb Z_2\rightarrow\SO(3)$ again, $s$ determines 
an element $\ul s\in\Omega^p(M,E)$. $\ul s$ is represented by a 
$\Omega^p\otimes\Bbb E_3$ \v Cech $0$--cochain glueing as
$$
\ul s_\alpha=\ul R_{\alpha\beta}\ul s_\beta
.\eqno(2.7)
$$
This correspondence establishes an isomorphism of the spaces 
$\Omega^p(M,\Ad P)$, $\Omega^p(M,E)$ of $p$--form sections of $\Ad P$, $E$.

Let $A\in\Conn(M,P)$ be a connection of $P$. The covariant derivative 
$D_As$ of $s\in\Omega^p(M,\Ad P)$ is 
$$
D_As=ds+[A,s]
.\eqno(2.8)
$$
$D_As\in\Omega^{p+1}(M,\Ad P)$. 
The isomorphisms $\Conn(M,P)\cong\Conn(M,E)$, $\Omega^p(M,\Ad P)\cong\Omega^p(M,E)$,
described above, map $D_As$ into the covariant derivative 
$\ul D_{\ul A}\ul s$ of $\ul s\in\Omega^p(M,E)$. Explicitly, 
$\ul D_{\ul A}\ul s$ is 
$$
\ul D_{\ul A}\ul s=d\ul s+\ul A\times\ul s
.\eqno(2.9)
$$
$\ul D_{\ul A}\ul s\in\Omega^{p+1}(M,E)$ as expected.

The gauge curvature of a connection $A\in \Conn(M,P)$ is 
$$
F_A=dA+\hbox{$1\over 2$}[A,A]
.\eqno(2.10)
$$
$F_A\in\Omega^2(M,\Ad P)$. 
Under the isomorphism $\Conn(M,P)\cong\Conn(M,E)$, 
$F_A$ is mapped into the gauge curvature $\ul F_{\ul A}$ of $\ul A$.
Explicitly, $\ul F_{\ul A}$ is 
$$
\ul F_{\ul A}=d\ul A+\hbox{$1\over 2$}\ul A\times \ul A
.\eqno(2.11)
$$
$\ul F_{\ul A}\in \Omega^2(M,E)$.

Let $U\in\Gau(M,P)$ be a gauge transformation of the principal bundle $P$. 
$U$ can be viewed as an $\ul{\SU}(2)$ \v Cech $0$--cochain
$\{U_\alpha\}$ satisfying
$$
U_\alpha=R_{\alpha\beta}U_\beta R_{\alpha\beta}{}^{-1}
.\eqno(2.12)
$$
By the isomorphism ${\sans H}:\SU(2)/\Bbb Z_2\rightarrow SO(3)$, 
$U$ yields a $\SO(3)$ valued endomorphism $\ul U\in \SO(M,E)$
of the vector bundle $E$ (an $\SO(3)$ endomorphism for short).
$\ul U$ is represented by an $\ul{\SO}(3)$ \v Cech $0$--cochain
$\{\ul U_\alpha\}$ satisfying
$$
\ul U_\alpha=\ul R_{\alpha\beta}\ul U_\beta\ul R_{\alpha\beta}{}^{-1}
.\eqno(2.13)
$$
This establishes an injective homomorphism of $\Gau(M,P)/\Bbb Z_2$ into $\SO(M,E)$,
where $\Bbb Z_2$ is realized as the sign group $\{\pm 1\}$. The homomorphism 
is not surjective in general.

A gauge transformation $U\in\Gau(M,P)$ acts on a connection 
$A\in \Conn(M,P)$ according to
$$
A^{\ul U}=\Ad U^{-1}A+U^{-1}dU
.\eqno(2.14)
$$
This action translates into one of the $\SO(3)$ endomorphism 
$\ul U\in\SO(M,E)$ corresponding to $U$ under the homomorphism
$\Gau(M,P)/\Bbb Z_2\rightarrow\SO(M,E)$ on the connection 
$\ul A\in\Conn(M,E)$ corresponding to $A$ under the isomorphism
$\Conn(M,P)\cong\Conn(M,E)$ given by 
$$
\ul A^{\ul U}=\ul U^{-1}\ul A+\ul\lambda_{\ul U}
.\eqno(2.15)
$$
where $\ul\lambda_{\ul U}$ is the $\Bbb E_3$ valued $1$--form defined by
$$
\star\,\ul\lambda_{\ul U}=-\ul U^{-1}d\ul U
.\eqno(2.16)
$$

A gauge transformation $U\in\Gau(M,P)$ acts on a section 
$s\in\Omega^p(M,\Ad P)$ as
$$
s^U=\Ad U^{-1}s
.\eqno(2.17)
$$
The homomorphisms $\Gau(M,P)/\Bbb Z_2\rightarrow\SO(M,E)$, 
$\Omega^p(M,\Ad P)\cong\Omega^p(M,E)$, described above, 
map $s^U$ into the result of the action of the $\SO(3)$ endomorphism $\ul U$
on $\ul s\in\Omega^p(M,E)$, $\ul s^{\ul U}$. The latter is given by
$$
\ul s^{\ul U}=\ul U^{-1}\ul s
.\eqno(2.18)
$$

All the isomorphisms established above will be repeatedly used in the following.

\titlebf {3. Cho structures}

In this section, we shall introduce the basic notion of Cho structure.
Cho structures provide a local formulation of Cho's original gauge theoretic
formalism \ref{23,24}.

\titlecps{Cho structures}

Let $M$ be a manifold. Let $P$ be a principal $\SU(2)$ bundle on $M$ and let 
$E$ be the vector bundle defined in (2.2). 

\defn {3.1} A {\it Cho structure} 
${\scri C}=(\{\ul n_i\},\{\ul A_i\},\{\ul T_{ij}\})$ of $E$ subordinated 
to the open cover $\{O_i\}$ of $M$ consists of the following elements.
\footnote{}{}\footnote{${}^2$}{Note that the cover $\{O_i\}$ 
is logically distinct from the cover $\{O_\alpha\}$ of sect. 2.}

\item{$i$)} 
A collection $\{\ul n_i\}$ of sections $\ul n_i\in\Omega^0(O_i,E)$ such that
$$
\ul n_i{}^2=1
.\eqno(3.1)
$$

\item{$ii$)} 
A collection $\{\ul A_i\}$ of connections $\ul A_i\in\Conn(O_i,E)$.

\item{$iii$)} 
A collection $\{\ul T_{ij}\}$ of sections $\ul T_{ij}\in\SO(O_{ij},E)$.

It is further assumed that these have the following properties.

\item {$a$)}
$\ul n_i$ is covariantly constant with respect to $\ul A_i$, 
$$
\ul D_{\ul A i}\ul n_i=0
.\eqno(3.2)
$$
where $D_{\ul A i}$ is the covariant derivative associated with $\ul A_i$
(cfr. eq. (2.9)).

\item {$b$)}
$\ul T_{ij}$ matches $\ul n_i$, $\ul n_j$,
$$
\ul n_i=\ul T_{ij}\ul n_j
.\eqno(3.3)
$$

\item {$c$)} 
The $\ul T_{ij}$ are normalized so that 
$$
\eqalignno{\vphantom{1\over 2}
&\ul T_{ii}=\ul 1,&(3.4a)\cr
\vphantom{1\over 2}
&\ul T_{ij}\ul T_{ji}=\ul 1.&(3.4b)\cr
}
$$

\noindent
We denote by ${\sans C}_E$ the set of Cho structures $\scri C$ of $E$
subordinated to some open cover of $M$.

\rema{3.2}
From (3.1), (3.2), it follows that a Cho structure ${\scri C}$ 
of $E$ describes a family 
of local $\SO(2)$ reduction of $E$ with compatible $\SO(2)$ reduced 
connections subordinated to the associated covering \ref{34}. 
We shall elaborate on this point in greater detail in sect. 5 below.

To compare Cho structures of $E$ subordinated to different open covers,
it is necessary to resort to refinement of the covers. 

\defprop{3.3}
Let ${\scri C}=(\{\ul n_i\},\{\ul A_i\},\{\ul T_{ij}\})$ be a Cho structure
of $E$ subordinated to the open cover $\{O_i\}$. Let 
$\{\bar O_{\bar\imath}\}$ be an open cover which is a refinement of 
$\{O_i\}$ and let $f$ be a refinement map (so that 
$\bar O_{\bar\imath}\subseteq O_{f(\bar\imath)}$). We set 
$$
\eqalignno{\vphantom{1\over 2}
\ul {\bar n}_{\bar\imath}&
=\ul n_{f(\bar\imath)}|_{\bar O_{\bar\imath}},&(3.5a)\cr
\vphantom{1\over 2}
\ul {\bar A}_{\bar\imath}&
=\ul A_{f(\bar\imath)}|_{\bar O_{\bar\imath}},&(3.5b)\cr
\vphantom{1\over 2}
\ul {\bar T}_{\bar\imath\bar\jmath}&
=\ul T_{f(\bar\imath)f(\bar\jmath)}|_{\bar O_{\bar\imath\bar\jmath}}.&(3.5c)\cr
}
$$
Then, $\bar {\scri C}=(\{\ul {\bar n}_{\bar\imath}\},
\{\ul {\bar A}_{\bar\imath}\},\{\ul {\bar T}_{\bar\imath\bar\jmath}\})$
is a Cho structure of $E$ subordinated to the cover 
$\{\bar O_{\bar\imath}\}$. We call $\bar {\scri C}$ the {\it refinement }
of ${\scri C}$ associated with the refinement $\{\bar O_{\bar\imath}\}$ of
$\{O_i\}$ and the refinement map $f$.

\proof The verification of (3.1)--(3.4) is straightforward. \hfill $\square$

\defprop{3.4}
Two Cho structures ${\scri C}=(\{\ul n_i\},\{\ul A_i\},\{\ul T_{ij}\})$, 
${\scri C}'=(\{\ul n'{}_i\},\{\ul A'{}_i\},\{\ul T'{}_{ij}\})$ 
of $E$ subordinated to the same open cover $\{O_i\}$
are called {\it equivalent} if they are related as
$$
\eqalignno{\vphantom{1\over 2}
\ul n'{}_i&=\ul U_i{}^{-1}\ul n_i,&(3.6a)\cr
\vphantom{1\over 2}
\ul A'{}_i&=\ul U_i{}^{-1}(\ul A_i-s_i\,\ul n_i) 
+\ul \lambda_{\ul U i},&(3.6b)\cr
\vphantom{1\over 2}
\ul T'{}_{ij}&
=\ul U_i{}^{-1}\ul T_{ij}\exp(-r_{ij}\star \ul n_j)\ul U_j,&(3.6c)\cr
}
$$
for some $\SO(E)$ \v Cech $0$--cochain $\{\ul U_i\}$ and some 
$\Omega^1$ \v Cech $0$--cochain $\{s_i\}$ and 
$\Omega^0$ \v Cech $1$--cochain $\{r_{ij}\}$
independent from the trivialization of $E$ used, where 
$\ul \lambda_{\ul U i}$ is defined by (2.16) with $\ul U$ replaced by
$\ul U_i$. More generally, two Cho
structures ${\scri C}=(\{\ul n_i\},\{\ul A_i\},\{\ul T_{ij}\})$, 
${\scri C}'=(\{\ul n'{}_{i'}\},\{\ul A'{}_{i'}\},\{\ul T'{}_{i'j'}\})$ 
of $E$ subordinated to the open covers
$\{O_i\}$, $\{O'{}_{i'}\}$ are called equivalent if there
is a common refinement $\{\bar O_{\bar\imath}\}$ of 
$\{O_i\}$, $\{O'{}_{i'}\}$ and refinement maps $f$, $f'$ such that the 
associated refinements $\bar {\scri C}=(\{\ul {\bar n}_{\bar\imath}\},
\{\ul {\bar A}_{\bar\imath}\},\{\ul {\bar T}_{\bar\imath\bar\jmath}\})$,
$\bar {\scri C}'=(\{\ul {\bar n}'{}_{\bar\imath}\},
\{\ul {\bar A}'{}_{\bar\imath}\},\{\ul {\bar T}'{}_{\bar\imath\bar\jmath}\})$
are equivalent in the restricted sense just defined. 
The above defines an equivalence relation on the set of Cho structures
of $E$. In the following, we denote by $[{\scri C}]$ the equivalence class of 
the Cho structure $\scri C$. 

\proof It is easy to see that (3.6) is consistent with (3.1)--(3.4). 
It is straightforward though tedious to verify the axioms of 
equivalence relation. \hfill $\square$

\rema {3.5}
The notion of equivalence defined in (3.6)
is more general than gauge equivalence, to which it reduces when
$s_i=0$ and $r_{ij}=0$. 

\rema {3.6} A Cho structure $\scri C$ is equivalent to all its refinements 
$\bar{\scri C}$.

\titlecps {Discussion}

The topological setting in which interesting Cho structures are defined has the 
following features. 
There are a manifold $M_0$, a principal $\SU(2)$ bundle $P_0$ on $M_0$ and a closed
submanifold $N_{\scri C}$ of $M_0$ depending on $\scri C$ such that 
$$
M=M_0\setminus N_{\scri C},
\eqno(3.7)
$$
$$
P=P_0|_M
\eqno(3.8)
$$
and that $\ul n_i$, $\ul A_i$ become singular on approaching 
$N_{\scri C}$, when $\partial O_i$ intersects $N$. That is, 
$N_{\scri C}$ is their defect manifold. $N_{\scri C}$ has codimension 3,
basically because the $\ul n_i$ are valued in the 3--dimensional vector
space $\Bbb E_3$. Further, there is a submanifold $D$ of $M_0$ bounded by 
$N_{\scri C}$, 
$$
D=\partial N_{\scri C},
\eqno(3.9)
$$
such that $O_{ij}\cap D=\emptyset$ for all $i,~j$ and
that the $\ul T_{ij}$ become singular when extended, if possible, 
beyond their domain $O_{ij}$ to a larger one intersecting $D$.
$D$ has codimension 2. $D$ is not uniquely defined by $\scri C$
and is a generalization of the well known Dirac string.

The physical origin of the setting just described has been illustrated in the 
introduction (see also \ref{19,20}).
In physical applications, $M_0$ is some $4$--dimensional space--time manifold, 
such as $\Bbb R^4$, $\Bbb S^4$, $\Bbb S^1\times\Bbb R^3$, $\Bbb T^4$,
$P_0$ is a principal $\SU(2)$ bundle defined by the $\SU(2)$ valued monodromy matrix 
functions of the gauge fields and $N_{\scri C}$ is a set of closed monopole world 
lines forming a knot in $M_0$. 

It is important to realize that the constructions worked out in this paper are fully 
general and do not require that $M$ and $P$ are of the form (3.7), (3.8) or that 
$\dim M=4$.

\titlecps{Examples}

We now illustrate the above analysis with a few examples.

\exa {3.7} {\it A small monopole loop in a single instanton background}

The authors of ref. \ref{35} found a solution of the differential maximal 
Abelian gauge in a single instanton background. Its defect manifold is a 
loop. When the radius of the loop is much smaller than the instanton size, 
an analytic expression is available. 

In this case, $M_0$ is the $4$--sphere $\Bbb S^4$ and $P_0$ is the trivial principal 
bundle $\Bbb S^4\times \SU(2)$. Here, we view $\Bbb S^4$ as the one point
compactification of $\Bbb R^4$, $\Bbb S^4=\Bbb R^4\cup\{\infty\}$.
We define coordinates $u,~v\in[0,+\infty]$, $\varphi,~\psi\in[0,2\pi[$ 
in $\Bbb R^4\cup\{\infty\}$ by
$$
x^1+\sqrt{-1}x^2=u\exp\big(\sqrt{-1}\varphi\big), \quad
x^3+\sqrt{-1}x^4=v\exp\big(\sqrt{-1}\psi\big).
\eqno(3.10)
$$
Note that these are ill defined at the planes $x^1=x^2=0$, $x^3=x^4=0$ and at 
infinity. Then, $M=\Bbb S^4\setminus N$, $P=(\Bbb S^4\setminus N)\times \SU(2)$ and 
$E=(\Bbb S^4\setminus N)\times \Bbb E_3$, where 
$$
N=\{x|x\in\Bbb R^4,~u=R_0,~v=0\},
\eqno(3.11)
$$
with $R_0>0$.

We set 
$$
\eqalignno{\vphantom{1\over 2}
O_1&=\{x|x\in\Bbb R^4\cup\{\infty\},~u<R_1\hbox{~and~}v<R_1\}\setminus N, &(3.12)\cr
\vphantom{1\over 2}
O_2&=\{x|x\in\Bbb R^4\cup\{\infty\},~u>R_2\hbox{~or~}v>R_2\}, &\cr
}
$$
where $R_1>R_2>R_0$. Note that $N\cap O_2=\emptyset$.
$\{O_1,O_2\}$ is an open covering of $M$.

Let $\alpha:\Bbb R^4\cup\{\infty\}\rightarrow[0,2\pi[$, 
$\beta:\Bbb R^4\cup\{\infty\}\rightarrow[0,\pi]$,
$\gamma:\Bbb R^4\cup\{\infty\}\rightarrow[0,2\pi[$ be given by
$$
\eqalignno{\vphantom{1\over 2}
\alpha&=\varphi-\psi \quad\hbox{mod $2\pi$}, &(3.13a)\cr
\vphantom{1\over 2}
\beta &=\tan^{-1}\Big({2uv\over u^2-v^2-R_0{}^2}\Big)
\quad\hbox{mod $\pi$}, &(3.13b)\cr
\vphantom{1\over 2}
\gamma&= \varphi+\psi \quad\hbox{mod $2\pi$}.&(3.13c)\cr
}
$$
$\alpha$, $\gamma$ are ill defined at the planes 
$u=0$, $v=0$, $\beta$ is ill defined at the hypersurfaces $u^2-v^2-R_0{}^2=0$.
Let $s_1$, $s_2$, $\exp\big(-\sqrt{-1}\chi_{12}\big)$ be
the local representatives of a connection $s$ and the transition function 
of some $\Bbb T$ principal bundle on $M$, respectively, so that 
$$
s_2-s_1=-d\chi_{12}
.\eqno(3.14)
$$
We set
$$
\eqalignno{\vphantom{1\over 2}
\ul n_1&=\big[\sin\beta(\cos\alpha\,\ul e_1+\sin\alpha\,\ul e_2)
+\cos\beta\,\ul e_3\big]\big|_{O_1}, &(3.15)\cr
\vphantom{1\over 2}
\ul n_2&=\ul e_3, &\cr
\vphantom{1\over 2}
\ul A_1&=\big[-d\alpha\,\ul e_3
-d\beta\big(-\sin\alpha\,\ul e_1+\cos\alpha\,\ul e_2\big)&(3.16)\cr
\vphantom{1\over 2}
&\hphantom{=}~
+(\cos\beta d\alpha-s_1)\big(\sin\beta(\cos\alpha\,\ul e_1+\sin\alpha\,\ul e_2)
+\cos\beta\,\ul e_3)\big]\big|_{O_1}, &\cr
\vphantom{1\over 2}
\ul A_2&=-s_2\ul e_3, &\cr
\vphantom{1\over 2}
\ul T_{12}&=\exp(-\alpha\star\ul e_3)\exp(-\beta \star\ul e_2)
\exp(-(\gamma+\chi_{12})\star\ul e_3)\big|_{O_{12}}, &(3.17)\cr
\vphantom{1\over 2}
\ul T_{21}&=\exp((\gamma+\chi_{12})\star\ul e_3)\exp(\beta \star\ul e_2)
\exp(\alpha\star\ul e_3)\big|_{O_{12}}, &\cr
\vphantom{1\over 2}
\ul T_{11}&=\ul T_{22}=\ul 1. &\cr
}
$$
It is straightforward to check that (3.15)--(3.17) define a Cho structure 
$\scri C$ of $E$ subordinated to $\{O_1,O_2\}$. Its defect manifold 
$N_{\scri C}$ is precisely $N$,
$$
N_{\scri C}=N
\eqno(3.18)
$$
\ref{35}. A Dirac sheet $D$ is the disk bounded by $N$ in the plane $v=0$.

\exa {3.8} {\it A Higgs field on a $4$--torus}

In this case, $M_0$ is the $4$--torus $\Bbb T^4$ and $P_0$ is the trivial 
principal bundle $\Bbb T^4\times \SU(2)$. The defect manifold $N$ is empty, 
$N=\emptyset$. Thus, $M=\Bbb T^4$, $P=\Bbb T^4\times \SU(2)$ and 
$E=\Bbb T^4\times \Bbb E_3$. 

We view $\Bbb T^4$, as the quotient of $\Bbb R^4$ by a lattice 
$\Lambda\subseteq \Bbb R^4$, $\Bbb T^4=\Bbb R^4/\Lambda$. 
Let $\{O_i\}$ an open covering of $\Bbb T^4$
such that, for each $i$, $O_i$ is a simply connected non empty open subset of 
$\Bbb T^4$. As is well known, for each $i$, there is a local coordinate 
$x_i$ of $\Bbb T^4$ defined on $O_i$ such that 
$$
\theta=x_i(\theta)+\Lambda,
\eqno(3.19)
$$
for $\theta\in O_i$. Further, when $O_{ij}\not=\emptyset$, there is 
$\xi_{ij}\in\Lambda$ such that
$$
x_i=x_j+\xi_{ij}\quad \hbox{on $O_{ij}$}
\eqno(3.20)
$$
The collection $\{\xi_{ij}\}$ is a $\Lambda$ valued \v Cech 
$1$--cocycle.

Let $c:\Bbb T^4\rightarrow\Bbb R^{4\vee}$ be a smooth function
and $\ul k_0,~\ul e_0\in \Bbb E_3$ be unit vectors.
Let $s_i$, $\exp\big(-\sqrt{-1}\chi_{ij}\big)$ be
the local representatives of a connection $s$ and the transition function 
of some $\Bbb T$ principal bundle on $M_0$, respectively, so that 
$$
s_j-s_i=-d\chi_{ij}
.\eqno(3.21)
$$
We set 
$$
\eqalignno{\vphantom{1\over 2}
\ul {\hat n}_i&=\exp\big(-\langle c, x_i\rangle\star \ul k_0\big)\ul e_0,
&(3.22)\cr
\vphantom{1\over 2}
\ul {\hat A}_i&=-\exp\big(-\langle c, x_i\rangle\star \ul k_0\big)
\big[d\langle c, x_i\rangle\ul e_0\times(\ul k_0\times\ul e_0)+s_i\ul e_0\big],
&(3.23)\cr
\vphantom{1\over 2}
\ul {\hat T}_{ij}&=\exp\big(-\langle c, x_i\rangle\star \ul k_0\big)
\exp\big(-\chi_{ij}\star \ul e_0\big)
\exp\big(\langle c, x_j\rangle\star \ul k_0\big)|_{O_{ij}}.&(3.24)\cr
}
$$
It is straightforward to check that (3.20)--(3.23) define a Cho structure 
$\hat{\scri C}$ of $E$ subordinated to $\{O_i\}$.
Its defect manifold $N_{\scri C}$ is empty
$$
N_{\scri C}=\emptyset
\eqno(3.25)
$$

\titlebf {4. The Deligne cohomology class $D_{\scri C}$}

To any equivalence class of Cho structures, there is associated a flat 
degree $3$ length $3$ Deligne cohomology class. In this section, we describe its 
construction. 

\titlecps{Construction of the Deligne cohomology class $D_{\scri C}$}

Let $M$, $P$ and $E$ be as in sect. 3.
Let ${\scri C}=(\{\ul n_i\},\{\ul A_i\},\{\ul T_{ij}\})$ be a
Cho structure of $E$ subordinated to the open cover $\{O_i\}$
(cfr. def. 3.1). 

\defprop {4.1} One has
$$
\ul A_i=\ul A_{\ul n i}-a_i\ul n_i,
\eqno(4.1)
$$
where
$$
\ul A_{\ul n i}=\ul A_0\cdot\ul n_i\,\ul n_i+d\ul n_i\times \ul n_i,
\eqno(4.2)
$$
$\ul A_0\in \Conn(M,E)$ is a fixed background connection of $E$ and 
$a_i\in\Omega^1(O_i)$. $\ul A_{\ul n i}\in\Conn(O_i, E)$ 
is a connection of $E$ on $O_i$ and $a_i$ does not depend on the 
trivialization of $E$ used. 

\proof This follows easily from (2.9), (3.2). \hfill $\square$

\rema {4.2} The connection $\ul A_i$ has the body fixed form of ref. \ref{21}.

\defprop {4.3} Let $\ul F_{\ul A i}$ be the curvature of $\ul A_i$ (cfr. eq. (2.11)).
There is a $2$--form $\eta_i\in\Omega^2(O_i)$ such that
$$
\ul F_{\ul A i}=-\eta_i\ul n_i.
\eqno(4.3)
$$
$\eta_i$ is given explicitly by 
$$
\eta_i=\hbox{$1\over 2$}\ul n_i\cdot d\ul n_i\times d\ul n_i
-d(\ul A_0\cdot\ul n_i)+da_i.
\eqno(4.4)
$$
$\eta_i$ does not depend on the trivialization of $E$ used. 

\proof From (3.2) and the Ricci identity $\ul D_{\ul Ai}\ul D_{\ul Ai}\ul n_i
=\ul F_{\ul A i}\times\ul n_i$, one finds that
$$
\ul F_{\ul A i}\times \ul n_i=0
.\eqno(4.5)
$$
It follows that $\ul F_{\ul A i}$ is of the form (4.3). The expression (4.4)
of $\eta_i$ follows readily from (2.11), (3.1), (4.1), (4.2). The last statement 
is obvious from (4.3). \hfill $\square$

\defprop {4.4} There is a $1$--form $\psi_{ij}\in\Omega^1(O_{ij})$ such that 
$$
\ul A_i=\ul T_{ij}(\ul A_j-\ul\zeta_{ij}+\psi_{ij}\ul n_j),
\eqno(4.6)
$$
where $\ul \zeta_{ij}$ is the $\Bbb E_3$ valued $1$--form defined by
$$
\star\,\ul \zeta_{ij}=-\ul T_{ij}{}^{-1}d\ul T_{ij}.
\eqno(4.7)
$$
$\psi_{ij}$  is given explicitly by 
$$
\psi_{ij}=\ul \zeta_{ij}\cdot\ul n_j+\ul A_0\cdot(\ul n_i-\ul n_j)-a_i+a_j.
\eqno(4.8)
$$
$\psi_{ij}$ does not depend on the trivialization of $E$ used.

\proof Combining (3.2), (3.3), one finds that 
$$
(\ul A_i-\ul T_{ij}(\ul A_j-\ul\zeta_{ij}))\times \ul T_{ij}\ul n_j=0.
\eqno(4.9)
$$
It follows that $\ul A_i$ is of the form (4.6) for some 
$\psi_{ij}\in\Omega^1(O_{ij})$. 
$\psi_{ij}$ can be computed explicitly from (3.1), (3.3), (4.1), (4.2).
The last statement is easily verified. \hfill $\square$ 

\defprop {4.5} There is a $0$--form $\phi_{ijk}\in\Omega^0(O_{ijk})$
such that
$$
\ul T_{ij}\ul T_{jk}\ul T_{ki}=\exp(-\phi_{ijk}\star \ul n_i)
.\eqno(4.10)
$$
$\phi_{ijk}$ does not depend on the trivialization of $E$ used. 

\proof The consistency of (3.3) requires that
$$
\ul n_i=\ul T_{ij}\ul T_{jk}\ul T_{ki}\ul n_i
.\eqno(4.11)
$$
From here it follows that there is $\phi_{ijk}\in\Omega^0(O_{ijk})$
satisfying (4.10). The last statement is obvious. \hfill $\square$ 

\defn {4.6} We set
$$
K_{ijkl}=\phi_{jkl}-\phi_{ikl}+\phi_{ijl}-\phi_{ijk}. 
\eqno(4.12)
$$

\rema {4.7} The $\phi_{ijk}$ are defined modulo $2\pi\Bbb Z$. 
So, we are allowed to redefine
$$
\phi_{ijk}\rightarrow\phi_{ijk}+ m_{ijk},
\eqno(4.13)
$$ 
where $m_{ijk}\in 2\pi\Bbb Z$, if we wish so.
To this, there corresponds an obvious redefinition of $K_{ijkl}$.

\defprop {4.8} $\{\eta_i\}$ is an $\Omega^2$ \v Cech $0$--cochain,
$\{\psi_{ij}\}$ is an $\Omega^1$ \v Cech $1$--cochain,
$\{\phi_{ijk}\}$ is an $\Omega^0$ \v Cech $2$--cochain
and $\{K_{ijkl}\}$ is a $2\pi\Bbb Z$ \v Cech $3$--cochain
(upon suitably fixing the indeterminacy (4.13)).
Further, one has 
$$
\eqalignno{\vphantom{1\over 2}
\delta\eta_{ij}&=d\psi_{ij},&(4.14a)\cr
\vphantom{1\over 2}
\delta\psi_{ijk}&=d\phi_{ijk},&(4.14b)\cr
\vphantom{1\over 2}
\delta\phi_{ijkl}&=K_{ijkl},&(4.14c)\cr
\vphantom{1\over 2}
\delta K_{ijklm}&=0,&(4.14d)\cr
}
$$
where $\delta$ is the \v Cech coboundary operator defined in (A.7). 
As a consequence, the sequence $(\{\eta_i\},\{\psi_{ij}\},\{\phi_{ijk}\},
\{K_{ijkl}\})$ is a length $3$ Deligne $3$--cocycle (cfr. app. A, eq. (A.4))
and, thus, it defines a degree $3$ length $3$ Deligne cohomology class 
$D_{\scri C}\in H^3(M,D(3)^\bullet)$. This class is unaffected by any redefinition of 
the form (4.13) and is thus associated with the Cho structure $\scri C$. 

\proof Using (3.4), (4.7), (4.10), it is a straightforward matter to verify that
$\{\eta_i\}$, $\{\psi_{ij}\}$, $\{\phi_{ijk}\}$ and $\{K_{ijkl}\}$ are 
Cech cochains of the stated types. The $2\pi\Bbb Z$ valuedness of 
$\{K_{ijkl}\}$ follows from the relation
$$
\exp(-\delta\phi_{ijkl}\star \ul n_i)=\ul 1,
\eqno(4.15)
$$
which can be shown by repeated application of (4.10). 
The proof of relations (4.14) is straightforward by observing that,
from (4.3), (4.6), 
$$
\delta\eta_{ij}=\ul F_{\ul A i}\cdot\ul n_i-\ul F_{\ul A j}\cdot\ul n_j,
\eqno(4.16)
$$
$$
\delta\psi_{ijk}=\ul \zeta_{jk}\cdot\ul n_k-\ul \zeta_{ik}\cdot\ul n_k
+\ul \zeta_{ij}\cdot\ul n_j
\eqno(4.17)
$$
and upon using (3.1), (3.3), (3.4), (4.6), (4.7), (4.10) and the identities
$$
\eqalignno{\vphantom{1\over 2}
\exp(-\varphi\star\ul n)&=\ul 1
+(1-\cos\varphi)(\star\ul n{})^2-\sin\varphi\star\ul n,
&(4.18)\cr
\vphantom{1\over 2}
\exp(\varphi\star\ul n)d\exp(-\varphi\star\ul n)
&=-d\varphi\star\ul n-\sin\varphi\star d\ul n
-(1-\cos\varphi)\star(d\ul n\times\ul n),&(4.19)\cr
}
$$
where $\varphi$ is a local $0$--form. The remaining statements are easily verified.
\hfill $\square$

Next, we shall study the dependence of $D_{\scri C}$ on ${\scri C}\in{\sans C}_E$.

\prop {4.9} Let ${\scri C}$ be a Cho structure of $E$ subordinated to the open 
cover $\{O_i\}$. Let $\{\bar O_{\bar\imath}\}$ be an open cover which is a refinement 
of $\{O_i\}$ and let $f$ be a refinement map. Let 
$\bar {\scri C}$ be the associated refinement of $\scri C$ (cfr. def. 3.3).
Then, 
$$
D_{\scri C}=D_{\bar{\scri C}}
\eqno(4.20)
$$

Thus, refinement of Cho structures is compatible with Deligne cohomology. 

\proof It is readily verified that 
$$
\eqalignno{\vphantom{1\over 2}
\bar\eta_{\bar\imath}&
=\eta_{f(\bar\imath)}|_{\bar O_{\bar\imath}},&(4.21a)\cr 
\vphantom{1\over 2}
\bar\psi_{\bar\imath\bar\jmath}&
=\psi_{f(\bar\imath)f(\bar\jmath)}|_{\bar O_{\bar\imath\bar\jmath}},&(4.21b)\cr
\vphantom{1\over 2}
\bar\phi_{\bar\imath\bar\jmath\bar k}&
=\phi_{f(\bar\imath)f(\bar\jmath)f(\bar k)}|_{\bar O_{\bar\imath
\bar\jmath\bar k}},&(4.21c)\cr
\vphantom{1\over 2}
\bar K_{\bar\imath\bar\jmath\bar k\bar l}&
=K_{f(\bar\imath)f(\bar\jmath)f(\bar k)f(\bar l)}|_{\bar O_{\bar\imath
\bar\jmath\bar k\bar l}},&(4.21d)\cr
}
$$
with obvious notation. This means that the Deligne $3$--cocycle corresponding to 
$\bar {\scri C}$ is the refinement of the Deligne $3$--cocycle corresponding to 
$\scri C$ associated with the refinement $\{\bar O_{\bar\imath}\}$ of
$\{O_i\}$ and the refinement map $f$. (4.20) follows immediately.
\hfill $\square$

\prop {4.10} Let ${\scri C}$, ${\scri C}'$ be equivalent Cho structures of $E$
(cfr. def. 3.4). Then,
$$
D_{\scri C}=D_{{\scri C}'}.
\eqno(4.22)
$$

Thus, the Deligne cohomology classes of equivalent Cho structures 
are equal. Since $D_{\scri C}$ depends on the Cho structure $\scri C$
only through its equivalence class $[{\scri C}]$, we shall use occasionally 
the notation $D_{[{\scri C}]}$. 

\proof Assume first that ${\scri C}$, ${\scri C}'$ are subordinated to the same 
open cover $\{O_i\}$. 
As ${\scri C}$, ${\scri C}'$ are equivalent, (3.6) holds.
Using (4.1), (4.4), (4.7), (4.8) and (4.10) and exploiting (4.19), 
it is straightforward to show that $\{s_i\}$ is a  
$\Omega^1$ \v Cech $0$--cochain, $\{r_{ij}\}$ is a  
$\Omega^0$ \v Cech $1$--cochain and that, 
for some $2\pi\Bbb Z$ \v Cech $2$--cochain $\{k_{ijk}\}$,
$$
\eqalignno{\vphantom{1\over 2}
\eta'{}_i&=\eta_i+ds_i ,&(4.23a)\cr 
\vphantom{1\over 2}
\psi'{}_{ij}&=\psi_{ij}+\delta s_{ij}+dr_{ij} ,&(4.23b)\cr
\vphantom{1\over 2}
\phi'{}_{ijk}&=\phi_{ijk}+\delta r_{ijk}+ k_{ijk} ,&(4.23c)\cr
\vphantom{1\over 2}
K'{}_{ijkl}&=K_{ijkl}+\delta k_{ijkl} .&(4.23d)\cr
}
$$
Therefore, the Deligne $3$--cocycles corresponding to $\scri C$, ${\scri C}'$ differ 
by a length $3$ Deligne $3$--coboundary (cfr. eq. (A.5)). It follows immediately that
$D_{\scri C}=D_{{\scri C}'}$. Next, let ${\scri C}$, ${\scri C}'$ 
be subordinated to distinct open covers $\{O_i\}$, $\{O'{}_{i'}\}$. 
Assume that ${\scri C}$, ${\scri C}'$ are equivalent.
Then, there is a common refinement $\{\bar O_{\bar\imath}\}$ of 
$\{O_i\}$, $\{O'{}_{i'}\}$ and refinement maps $f$, $f'$ such that the 
associated refinements $\bar {\scri C}$, $\bar {\scri C}'$
are equivalent. By the result just shown, $D_{\bar{\scri C}}=D_{\bar{\scri C}'}$. 
On the other hand, by (4.20), $D_{\scri C}=D_{\bar{\scri C}}$, 
$D_{{\scri C}'}=D_{\bar{\scri C}'}$. (4.22) follows. \hfill $\square$

\rema {4.11} Since a Cho structure
$\scri C$ is equivalent to anyone of its refinements $\bar{\scri C}$, 
(4.20) is a particular case of (4.22).

\prop {4.12} For any Cho structure $\scri C$, the Deligne class $D_{\scri C}$ 
is flat.

\proof From (3.2), (4.3) and the Bianchi identity $\ul D_{\ul Ai}\ul F_{\ul Ai}=0$, 
one verifies that
$$
-d\eta_i\ul n_i=0
\eqno(4.24)
$$
and, so, the $2$--form $\eta_i$ is closed
$$
d\eta_i=0.
\eqno(4.25)
$$
This shows that $D_{\scri C}$ is flat as stated. \hfill $\square$

\titlecps{The main theorem on $D_{\scri C}$}

All the above results are summarized in the following theorem.

\theo {4.13} There is a well defined natural map that associates
with any equivalence class of Cho structures $[{\scri C}]$
a flat degree $3$ length $3$ Deligne cohomology class 
$D_{[{\scri C}]}\in H^3(M,D(3)^\bullet)$. 

\rema {4.14}
As discussed in app. A, with $D_{\scri C}$ there is associated 
an isomorphism class of Hermitian gerbes with Hermitian connective 
structure and curving. By prop. 4.12, this class is flat.

This was to be expected in the present finite dimensional 
context (see ref. \ref{32} for a discussion of this matter).
Later, in sect. 6, we shall show that the class actually trivial.

\titlecps{Examples }

\exa {4.15} {\it A small monopole loop in a single instanton background}

This example was illustrated in sect. 3. 
It is straightforward to compute the Deligne $3$--cocycle associated with 
the Cho structure $\scri C$ (3.15)--(3.17).
Its only non vanishing components are 
$$
\eta_1=\big[-\sin\beta d\alpha d\beta\big]\big|_{O_1}+ds_1,\quad \eta_2=ds_2,
\eqno(4.26)
$$
$$
\psi_{12}=-\psi_{21}=\big[d\gamma+\cos\beta d\alpha\big]\big|_{O_{12}}. 
$$

\exa {4.16}  {\it A Higgs field on a $4$--torus}

Also this example was illustrated in sect. 3.
It is simple to compute the Deligne $3$--cocycle associated associated with 
the Cho structure ${\scri C}$ defined (3.22)--(3.24)
$$
\eqalignno{\vphantom{1\over 2}
\eta_i&=t|_{O_i},&(4.27)\cr 
\vphantom{1\over 2}
\psi_{ij}&=0,&\cr
\vphantom{1\over 2}
\phi_{ijk}&=0,&\cr
\vphantom{1\over 2}
K_{ijkl}&=0,&\cr
}
$$
where $t\in\Omega^2_{c2\pi\Bbb Z}(M)$ is a closed $2$--form with 
periods in $2\pi\Bbb Z$ defined by $t|_{O_i}=ds_i$.  

\titlebf {5. Fine and almost fine Cho structures and the Deligne class $D_{\scri C}$}

In general, the local data of a Cho structure cannot be assembled in a global 
structure, but they do when the structure is fine. 

\titlecps{Fine Cho structures}

Let $M$, $P$ and $E$ be as in sect. 3.

\defn {5.1}  A Cho structure ${\scri C}=(\{\ul n_i\},\{\ul A_i\},\{\ul T_{ij}\})$ of 
$E$ subordinated to the open cover $\{O_i\}$ (cfr. def. 3.1) is said {\it fine} fine 
if 
$$
\ul T_{ij}\ul T_{jk}\ul T_{ki}=\ul 1
\eqno(5.1)
$$
and 
$$
\ul A_i=\ul T_{ij}(\ul A_j-\ul\zeta_{ij})
\eqno(5.2)
$$
(cfr. eqs. (4.6), (4.7). We denote by ${\sans C}_E{}^*$ the set of fine Cho structures 
of $E$ subordinated to some open cover of $M$.

\rema{5.2}
Let ${\scri C}$ be a fine Cho structure. If $\bar{\scri C}$ is a Cho structure
refining ${\scri C}$ (cfr. def. 3.3), then $\bar{\scri C}$, also, is fine. 
If ${\scri C}'$ is a Cho structure equivalent to ${\scri C}$ (cfr. def. 3.4), 
then ${\scri C}'$ is not fine in general. 

When dealing with fine Cho structures, one thus needs a sharper notion of 
equivalence compatible with fineness.

\defprop{5.3}
Two fine Cho structures  ${\scri C}=(\{\ul n_i\},\{\ul A_i\}$,
$\{\ul T_{ij}\})$, ${\scri C}'=(\{\ul n'{}_i\},
\{\ul A'{}_i\},\{\ul T'{}_{ij}\})$ of $E$ subordinated 
to the same open cover $\{O_i\}$ are said {\it finely 
equivalent}, if 
$$
\eqalignno{\vphantom{1\over 2}
\ul n'{}_i&=\ul U_i{}^{-1}\ul { n}_i,&(5.3a)\cr
\vphantom{1\over 2}
\ul A'{}_i&=\ul U_i{}^{-1}\ul A_i+\ul \lambda_{\ul U i},&(5.3b)\cr
\vphantom{1\over 2}
\ul T'{}_{ij}&=\ul U_i{}^{-1}\ul T_{ij}\ul U_j,&(5.3c)\cr
}
$$
for some $\SO(E)$ \v Cech $0$--cochain $\{\ul U_i\}$.
It is readily checked that these relations are compatible with
the fineness of ${\scri C}$, ${\scri C}'$. 
More generally, two fine Cho structures
${\scri C}=(\{\ul n_i\},\{\ul A_i\},\{\ul T_{ij}\})$, 
${\scri C}'=(\{\ul n'{}_{i'}\},\{\ul A'{}_{i'}\},\{\ul T'{}_{i'j'}\})$ 
of $E$ subordinated to the open covers
$\{O_i\}$, $\{O'{}_{i'}\}$ are called finely equivalent if there
is a common refinement $\{\bar O_{\bar\imath}\}$ of 
$\{O_i\}$, $\{O'{}_{i'}\}$ and refinement maps $f$, $f'$ such that the 
associated fine refinements $\bar{\scri C}
=(\{\ul {\bar n}_{\bar\imath}\},\{\ul {\bar A}_{\bar\imath}\},
\{\ul {\bar T}_{\bar\imath\bar\jmath}\})$,
$\bar {\scri C}{}'=(\{\ul {\bar n}'{}_{\bar\imath}\},
\{\ul {\bar A}'{}_{\bar\imath}\},
\{\ul {\bar T}'{}_{\bar\imath\bar\jmath}\})$
are finely equivalent in the restricted sense just defined. 
As suggested by the name, the above defines indeed an equivalence 
relation on the set of fine Cho structures. Below, we shall denote by 
$\langle{\scri C}\rangle$ the fine equivalence class of a fine Cho structure 
${\scri C}$. 

\proof These statements are straightforwardly verified. \hfill $\square$

\rema{5.4} Fine equivalence implies ordinary equivalence as defined in def. 3.4 above.

\defprop{5.5} A fine Cho structure $\scri C$ defines an $\SO(3)$ vector bundle 
$E_{\scri C}$, a connection $\ul A_{\scri C}\in\Conn(M,E_{\scri C})$
and a section $\ul n_{\scri C}\in\Omega^0(M,E_{\scri C})$ such that 
$$
\ul n_{\scri C}{}^2=1,
\eqno(5.4)
$$
$$
\ul D_{\ul A_{\scri C}}\ul n_{\scri C}=0.
\eqno(5.5)
$$
$E_{\scri C}$ decomposes as
$$
E_{\scri C}=\hat E_{\scri C}\oplus C_{\scri C}.
\eqno(5.6)
$$
Here, $\hat E_{\scri C}$, $C_{\scri C}$ are the vector subbundles of 
$E_{\scri C}$ of local sections $\ul s$ of $E_{\scri C}$ such that 
$\ul s\cdot \ul n_{\scri C}=0$, $\ul s\times\ul n_{\scri C}=0$, respectively. 
$\hat E_{\scri C}$ is an $\SO(2)$ vector bundle.
$C_{\scri C}$ is isomorphic to the trivial bundle $\Bbb R\times M$. 
The complexification $\hat E_{\scri C}{}^{\Bbb C}$ of  
$\hat E_{\scri C}$ decomposes as 
$$
\hat E_{\scri C}{}^{\Bbb C}=e_{\scri C}\oplus e_{\scri C}{}^{-1},
\eqno(5.7)
$$
where $e_{\scri C}$ is a Hermitian line bundle. 
The connection $\ul A_{\scri C}$ induces a Hermitian connection 
$a_{\scri C}\in\Conn(M,e_{\scri C})$ and the trivial connection on 
$C_{\scri C}$. The curvature $f_{a_{\scri C}}=da_{\scri C}$ of $a_{\scri C}$
is given by
$$
f_{a_{\scri C}}=-\ul F_{\ul A_{\scri C}}\cdot \ul n_{\scri C}.
\eqno(5.8)
$$
The Hermitian line bundle with Hermitian connection $(e_{\scri C},a_{\scri C})$
determines the $\SO(3)$ vector bundle with connection $(E_{\scri C},\ul A_{\scri C})$ 
up to isomorphism. 
Finely equivalent fine Cho structures ${\scri C}$, ${\scri C}'$ yield isomorphic 
vector bundles with connections $(E_{\scri C},\ul A_{\scri C})$,
$(E_{{\scri C}'},\ul A_{{\scri C}'})$.
The line bundle with connection $(e_{\scri C},a_{\scri C})$
depends only on the fine equivalence class $\langle{\scri C}\rangle$ 
of the fine Cho structure ${\scri C}$.

For this reason, we shall use occasionally the notation
$(e_{\langle{\scri C}\rangle},a_{\langle{\scri C}\rangle})$.

\proof Let ${\scri C}=(\{\ul n_i\},\{\ul A_i\},\{\ul T_{ij}\})$.
(5.1) implies that there is an $\SO(3)$ vector
bundle $E_{\scri C}$ obtained from the restrictions 
$E|_{O_i}$ by pasting through the $\SO(3)$ endomorphisms 
$\ul T_{ij}\in\SO(O_{ij}, E|_{O_{ij}})$. (5.1), (5.2)
imply in turn that the $\Conn(E)$ 0--cochain $\{\ul A_i\}$ 
defines a connection $\ul A_{\scri C}\in\Conn(M,E_{\scri C})$.  Then, 
the $\Omega^0(E)$ 0--cochain $\{\ul n_i\}$ yields a section 
$\ul n_{\scri C}\in\Omega^0(M,E_{\scri C})$ satisfying (5.4), (5.5). 
The decomposition of $E_{\scri C}$ follows straightforwardly from well known 
results about $\SO(2)$ reductions of $\SO(3)$ vector bundles \ref{34}. 
The remaining statements are easily proved. 
\hfill $\square$

Before proceeding to the mathematical study of fine Cho structures, 
it is necessary to realize the physical origin of this notion. 
Assume that $\dim M=4$. In the body fixed frame of \ref{21}, 
the data $E_{\scri C}$, $\ul A_{\scri C}$, $\ul n_{\scri C}$
associated with a fine Cho structure $\scri C$
are precisely those describing a monopole configuration.
The 1--dimensional defect manifold $N_{\scri C}$ of $\scri C$
(cfr. the discussion of sect. 3) is the set of the monopole world lines.
The integral $-{1\over 4\pi}\int_\Sigma \ul F_{\ul A_{\scri C}}
\cdot\ul n_{\scri C}$, where $\Sigma$ is any $2$--cycle of $M$, 
is the monopole charge enclosed by $\Sigma$.
These aspects will be discussed in sect. 7 in greater detail. 

\defprop{5.6} With any fine Cho structure $\scri C$, there is associated 
a degree $2$ length $2$ Deligne cohomology class $I_{\scri C}\in H^2(M,D(2)^\bullet)$. 
Its curvature $f_{\scri C}\in\Omega^2_{c2\pi\Bbb Z}(M)$ is 
a closed $2$--form with periods in $2\pi\Bbb Z$ called the 
{\it Abelian curvature} of $\scri C$. $f_{\scri C}$ is given by 
$$
f_{\scri C}=f_{a_{\scri C}}.
\eqno(5.9)
$$
$I_{\scri C}$, $f_{\scri C}$ depend only on the fine equivalence class 
$\langle{\scri C}\rangle$ of the fine Cho structure ${\scri C}$.

Again, we shall sometimes use the notation $I_{\langle{\scri C}\rangle}$,
$f_{\langle{\scri C}\rangle}$.

\proof By prop. 5.5, the fine Cho structure $\scri C$
determines a Hermitian line bundle with Hermitian connection 
$(e_{\scri C},a_{\scri C})$. 
As is well know, the isomorphism class of the latter
is represented by a degree $2$ length $2$ Deligne cohomology class
$I_{\scri C}\in H^2(M,D(2)^\bullet)$ with curvature $f_{\scri C}=f_{a_{\scri C}}
\in\Omega^2_{c2\pi\Bbb Z}(M)$ (cfr. app. A). The remaining statement  follows
easily from the last part of prop. 5.5. \hfill $\square$

\prop{5.7} If $\scri C$ is a fine Cho structure, then 
$$
D_{\scri C}=0
\eqno(5.10)
$$
(cfr. def. 4.8)

\proof Let ${\scri C}=(\{\ul n_i\},\{\ul A_i\},\{\ul T_{ij}\})$ be a 
fine Cho structure. We assume first that the open covering $\{O_i\}$ 
underlying ${\scri C}$ is good \ref{33}. 
Combining (4.3), (4.6), (4.10) and (5.1), (5.2), (5.8), one finds that 
the associated degree 3 length 3 Deligne cocycle 
$(\{\eta_i\},\{\psi_{ij}\},\{\phi_{ijk}\}$, $\{K_{ijkl}\})$ 
(cfr. prop. 4.8) is given by 
$$
\eqalignno{\vphantom{1\over 2}
\eta_i&=f_{a_{\scri C}}|_{O_i},&(5.11a)\cr 
\vphantom{1\over 2}
\psi_{ij}&=0,&(5.11b)\cr
\vphantom{1\over 2}
\phi_{ijk}&=m_{ijk},&(5.11c)\cr
\vphantom{1\over 2}
K_{ijkl}&=\delta m_{ijkl},&(5.11d)\cr
}
$$
where $\{m_{ijk}\}$ is some $2\pi\Bbb Z$ \v Cech 2--cochain. 
Here, $f_{a_{\scri C}}\in\Omega^2_{c2\pi\Bbb Z}(M)$. Since the cover
$\{O_i\}$ is good, this suffices to show that our Deligne cocycle is a coboundary 
(cfr. eq. (A.5)) \ref{32}. (5.10) follows. This result holds also when the cover 
$\{O_i\}$ is not good. Indeed, every open cover $\{O_i\}$ admits an open 
refinement $\{\bar O_{\bar\imath}\}$ which is good \ref{33}.
Pick a refinement map $f$. Let $\bar{\scri C}$ be the associated refinement. 
By rem. 5.2, $\bar{\scri C}$ is fine. Since the cover $\{\bar O_{\bar\imath}\}$
is good, $D_{\bar{\scri C}}=0$, by the result just proved. 
By prop. 4.10, eq. (4.22), one has then $D_{\scri C}=D_{\bar{\scri C}}=0$. 
Thus, (5.10) holds in general. \hfill $\square$

\titlecps{Almost fine Cho structures}

The Cho structures which show up in physical applications are seldom fine.
The following notion is therefore useful.

\defn{5.8} A Cho structure ${\scri C}$
of $E$ is said {\it almost fine} if it is equivalent to a fine Cho 
structure. 

To anticipate, in sect. 6, by exploiting local diagonalizability, 
we shall see that, actually, every Cho structure is almost fine.
However, this fact is not obvious {\it a priori}.

\rema{5.9} A Cho structure equivalent 
to an almost fine Cho structure is almost fine. In particular, 
a fine Cho structure is also almost fine.

\prop{5.10} A Cho structure $\scri C$ is almost fine if and only if 
$$
D_{\scri C}=0
.\eqno(5.12)
$$

\proof 
If ${\scri C}$ is an almost fine Cho structure, then $\scri C$ is equivalent
to a fine Cho structure ${\scri C}'$. By prop. 4.10, eq. (4.22), 
and prop. 5.7, eq. (5.10), $D_{\scri C}=D_{{\scri C}'}=0$. 
Hence, (5.12) holds. 

Conversely, let ${\scri C}=(\{\ul n_i\},\{\ul A_i\},\{\ul T_{ij}\})$ 
be a Cho structure satisfying (5.12). 
We assume first that the open covering $\{O_i\}$ underlying ${\scri C}$ 
is good. Then, the length $3$ Deligne $3$--cocycle 
$(\{\eta_i\},\{\psi_{ij}\},\{\phi_{ijk}\},\{K_{ijkl}\})$
associated with $\scri C$ is a coboundary. Therefore
$$
\eqalignno{\vphantom{1\over 2}
\eta_i&=-ds_i,&(5.13a)\cr
\vphantom{1\over 2}
\psi_{ij}&=-\delta s_{ij}-dr_{ij},&(5.13b)\cr
\vphantom{1\over 2}
\phi_{ijk}&=-\delta r_{ijk}-k_{ijk},&(5.13c)\cr
\vphantom{1\over 2}
K_{ijkl}&=-\delta k_{ijkl}.&(5.13d)\cr
}
$$
where $\{s_i\}$ is an $\Omega^1$ \v Cech $0$--cochain,
$\{r_{ij}\}$ is an $\Omega^0$ \v Cech $1$--cochain and
$\{k_{ijk}\}$ is a $2\pi\Bbb Z$ \v Cech $2$--cochain, by eq. (A.5).
Then, the sequence ${\scri C}'=(\{\ul n'{}_i\},\{\ul A'{}_i\},\{\ul T'{}_{ij}\})$ 
defined by 
$$
\eqalignno{\vphantom{1\over 2}
\ul n'{}_i&=\ul n_i,&(5.14a)\cr
\vphantom{1\over 2}
\ul A'{}_i&=\ul A_i-s_i\ul n_i,&(5.14b)\cr
\vphantom{1\over 2}
\ul T'{}_{ij}&=\ul T_{ij}\exp(-r_{ij}\star \ul n_j),&(5.14c)\cr
}
$$
is a Cho structure of $E$ subordinated to the cover $\{O_i\}$ 
equivalent to $\scri C$, by def. 3.4, eq. (3.6). By eq. (4.23), by construction, 
the associated length $3$ Deligne $3$--cocycle 
$(\{\eta'{}_i\},\{\psi'{}_{ij}\},\{\phi'{}_{ijk}\}$, $\{K'{}_{ijkl}\})$ 
is of the form
$$
\eqalignno{\vphantom{1\over 2}
\eta'{}_i&=0,&(5.15a)\cr 
\vphantom{1\over 2}
\psi'{}_{ij}&=0,&(5.15b)\cr
\vphantom{1\over 2}
\phi'{}_{ijk}&=m'{}_{ijk},&(5.15c)\cr
\vphantom{1\over 2}
K'{}_{ijkl}&=\delta m'{}_{ijkl},&(5.15d)\cr
}
$$
where $\{m'{}_{ijk}\}$ is a $2\pi\Bbb Z$ \v Cech $2$--cochain.
Thus, ${\scri C}'$ is fine. Consequently, 
$\scri C$ is almost fine. This result holds also when the cover 
$\{O_i\}$ is not good. Indeed, every open cover $\{O_i\}$ admits an open 
refinement $\{\bar O_{\bar\imath}\}$ which is good. 
Pick a refinement map $f$. Let $\bar{\scri C}$ be the associated refinement. 
By rem. 3.6, prop. 4.10, eq. (4.22), and (5.12),
$D_{\bar{\scri C}}=D_{\scri C}=0$.  
So, $\bar{\scri C}$ is a Cho structure subordinated to the good open cover 
$\{\bar O_{\bar\imath}\}$ satisfying (5.12). Thus, $\bar{\scri C}$ is almost 
fine, as just shown. On the other hand, $\scri C$ is equivalent to 
$\bar{\scri C}$. Thus, by rem. 5.9, $\scri C$ is almost fine as well.
\hfill $\square$

\rema{5.11} 
When $\scri C$ is almost fine, then $\scri C$ is equivalent to infinitely many 
fine Cho structures ${\scri C}'$.
It is fairly obvious that equivalent almost fine Cho structures $\scri C$
are characterized by the same set of fine Cho structures ${\scri C}'$.

\defn{5.12} We denote by ${\sans F}_{\scri C}$ the set of fine equivalence 
classes of fine Cho structures ${\scri C}'$ equivalent to an almost fine Cho 
structure $\scri C$. 

By rem. 5.11, we shall also use the notation ${\sans F}_{[{\scri C}]}$. 

\prop{5.13}
Let $\scri C$ be an almost fine Cho structure. Then, there exists a 
bijection $H^2(M,D(2)^\bullet)\cong{\sans F}_{\scri C}$.

\proof Let ${\scri C}=(\{\ul n_i\},\{\ul A_i\},\{\ul T_{ij}\})$.
By rem. 5.2 and rem. 5.11, we can assume without loss of generality that 
$\scri C$ itself is fine and that the open cover $\{O_i\}$ associated with 
$\scri C$ is good.  The degree 3 length 3 Deligne cocycle 
$(\{\eta_i\},\{\psi_{ij}\},\{\phi_{ijk}\},\{K_{ijkl}\})$
associated with $\scri C$ is thus of the form (5.11).

Let $J\in H^2(M,D(2)^\bullet)$. Let $(\{s_i\},\{-r_{ij}\},\{k_{ijk}\})$ be a 
degree 2 length 2 Deligne cocycle representing $J$. Therefore, one has
$$
\eqalignno{
\vphantom{1\over 2}
\delta s_{ij}+dr_{ij}&=0,&(5.16a)\cr
\vphantom{1\over 2}
\delta r_{ijk}+k_{ijk}&=0,&(5.16b)\cr
\vphantom{1\over 2}
\delta k_{ijkl}&=0&(5.16c)\cr
}
$$
by eq. (A.4). The sequence 
${\scri C}'=(\{\ul n'{}_i\},\{\ul A'{}_i\},\{\ul T'{}_{ij}\})$ 
defined by (5.14) is a Cho structure of $E$ equivalent to $\scri C$.
By using (4.23), it is readily checked that the associated Deligne cocycle
$(\{\eta'{}_i\},\{\psi'{}_{ij}\},\{\phi'{}_{ijk}\},\{K'{}_{ijkl}\})$
is also of the form (5.11). Hence, the Cho structure ${\scri C}'$ is
fine. If we replace the Deligne cocycle $(\{s_i\},\{-r_{ij}\},\{k_{ijk}\})$ by
a cohomologous Deligne cocycle $(\{s'{}_i\},\{-r'{}_{ij}\},\{k'{}_{ijk}\})$,
then the fine Cho structure 
${\scri C}'=(\{\ul n'{}_i\},\{\ul A'{}_i\},\{\ul T'{}_{ij}\})$ 
gets replaced by a finely equivalent fine Cho Structure
${\scri C}''=(\{\ul n''{}_i\},\{\ul A''{}_i\},\{\ul T''{}_{ij}\})$.
Indeed, by eq. (A.5), one has 
$$
\eqalignno{\vphantom{1\over 2}
s'_i&=s_i+dp_i,&(5.17a)\cr
\vphantom{1\over 2}
r'_{ij}&=r_{ij}-\delta p_{ij}-l_{ij},&(5.17b)\cr
\vphantom{1\over 2}
k'_{ijk}&=k_{ijk}+\delta l_{ijk},&(5.17c)\cr
}
$$
where $\{p_i\}$ is a $\Omega^0$ \v Cech $0$--cochain
and $\{l_{ij}\}$ is a $2\pi\Bbb Z$ \v Cech $1$--cochain.
Using (4.18), (4.19), it is straightforward to check that 
${\scri C}'$, ${\scri C}''$ are related as in (5.3)
with $\ul U_i=\exp(p_i\star\ul n_i)$ (after an obvious notational change).
${\scri C}'$, ${\scri C}''$ are thus finely equivalent fine Cho 
structures. Therefore, we have a well defined map
${\sans j}_{\scri C}:H^2(M,D(2)^\bullet)
\rightarrow {\sans F}_{\scri C}$
given by ${\sans j}_{\scri C}(J)=\langle{\scri C}'\rangle$.
It is easy to see that ${\sans j}_{\scri C}$
does not depend on the good open cover $\{O_i\}$. 

Next, we shall show that ${\sans j}_{\scri C}$ is a bijection.
We shall do so by proving that ${\sans j}_{\scri C}$ is injective 
and surjective. Let $J^{(a)}\in H^2(M,D(2)^\bullet)$, $a=1,~2$, such that 
${\sans j}_{\scri C}(J^{(1)})
={\sans j}_{\scri C}(J^{(2)})$.  Pick 2 length 2 Deligne cocycles
$(\{s^{(a)}{}_i\},\{-r^{(a)}{}_{ij}\},\{k^{(a)}{}_{ijk}\})$ representing $J^{(a)}$
and consider the representatives
${\scri C}^{(a)}=(\{\ul n^{(a)}{}_i\},\{\ul A^{(a)}{}_i\},\{\ul T^{(a)}{}_{ij}\})$ 
of ${\sans j}_{\scri C}(J^{(a)})$ given by (5.14), as defined 
in the previous paragraph. By assumption,  the fine Cho structures 
${\scri C}^{(1)}$, ${\scri C}^{(2)}$ are finely equivalent. So, 
they are related as in (5.3) for some $\SO(E)$ 0--cochain $\{\ul U_i\}$
(after an obvious notational change). 
On the other hand, for the representatives ${\scri C}^{(a)}$, 
$\ul n^{(a)}{}_i=\ul n_i$ so that $U_i=\exp(p_i\star\ul n_i)$ for an  
$\Omega^0$ \v Cech $0$--cochain $\{p_i\}$.
Using (4.18), (4.19), it is straightforward to check that the Deligne cocycles
$(\{s^{(a)}{}_i\},\{-r^{(a)}{}_{ij}\},\{k^{(a)}{}_{ijk}\})$ are related as in
(5.17) for some $2\pi\Bbb Z$ \v Cech $1$--cochain $\{l_{ij}\}$ 
(after another obvious notational change), so that they are cohomologous.
Hence, $J^{(1)}=J^{(2)}$. This shows that 
${\sans j}_{\scri C}$ is injective.
Next, let ${\scri C}'$ a fine Cho structure equivalent to $\scri C$.
Let us show that $\langle{\scri C}'\rangle$ is contained in the range of 
${\sans j}_{\scri C}$. By rem. 5.2 and for reasons explained 
in the previous paragraph, we can assume that $\scri C$, ${\scri C}'$ 
are subordinated to the same open cover. Further, we are free to replace ${\scri C}'$
be any other finely equivalent Cho structure subordinated to that cover.
Let ${\scri C}'=(\{\ul n'{}_i\},\{\ul A'{}_i\},\{\ul T'{}_{ij}\})$.
For the reasons just explained and the equivalence of $\scri C$, ${\scri C}'$, we  
can assume that ${\scri C}'$ is given by (5.14) for some $\Omega^1$ 
\v Cech $0$--cochain $\{s_i\}$ and $\Omega^0$ \v Cech $1$--cochain $\{r_{ij}\}$.
On the other hand, the Deligne cocycles 
$(\{\eta_i\},\{\psi_{ij}\},\{\phi_{ijk}\},\{K_{ijkl}\})$, 
$(\{\eta'{}_i\},\{\psi'{}_{ij}\},\{\phi'{}_{ijk}\},\{K'{}_{ijkl}\})$
are of the form (5.11) and they are related as in (4.23)
for some $2\pi\Bbb Z$ \v Cech $2$--cochain $\{k_{ijk}\}$.
It follows that, after perhaps a redefinition of $\{k_{ijk}\}$, 
the sequence $(\{s_i\},\{-r_{ij}\},\{k_{ijk}\})$ is a degree 2 length 2 
Deligne cocycle. This defines a class $J\in H^2(M,D(2)^\bullet)$. 
From the construction, it is obvious that 
${\sans j}_{\scri C}(J)=\langle{\scri C}'\rangle$.
Hence, ${\sans j}_{\scri C}$ is surjective 
\hfill $\square$

\rema{5.14} The bijection indicated in theor. (5.12) is not canonical,
since it depends on a choice of a reference fine structure in the class
$[{\scri C}]$.

The relevance of the classification result of prop. 5.13
will become clear in sect. 7.

\titlecps{Examples}

\exa{5.15} {\it A small monopole loop in a single instanton background}

This example was illustrated in sect. 3. From (4.26), it appears that 
the Cho structure $\scri C$ defined in (3.15)--(3.17) is not fine.
Later, we shall see that $\scri C$ is almost fine. 

\exa{5.16} {\it A Higgs field on a $4$--torus}

Also this example was illustrated in sect. 3. From (4.27), it is evident that 
the Cho structure $\hat{\scri C}$ defined (3.22)--(3.24) is fine
and, thus, also almost fine.

\titlebf{6. Diagonalizability and almost fineness}

In this section, we introduce a new type of Cho structures, the 
diagonalizable ones. In due course, we shall show that 
every diagonalizable Cho structure is almost fine
(cfr. def. 5.8) and that every Cho structure is diagonalizable. In this way, 
we shall conclude that every Cho structure is almost fine. 

\titlecps{Diagonalizable Cho structures}

Let $M$, $P$ and $E$ be as in sect. 3.

\defn{6.1}
A Cho structure ${\scri C}=(\{\ul n_i\},\{\ul A_i\},\{\ul T_{ij}\})$
of $E$ subordinated to the open cover $\{O_i\}$ is said 
{\it diagonalizable} if it is of the form
$$
\eqalignno{\vphantom{1\over 2}
\ul n_i&=\ul S_i\ul n_0,&(6.1a)\cr
\vphantom{1\over 2}
\ul A_i&=-\ul S_i(\theta_i\ul n_0+\ul \lambda_{\ul S i}),&(6.1b)\cr
\vphantom{1\over 2}
\ul T_{ij}&=\ul S_i
\exp(- \omega_{ij}\star\ul n_0)\ul S_j{}^{-1},&(6.1c)\cr
}
$$
where $\ul n_0\in \Bbb E_3$ with $\ul n_0{}^2=1$, $\{\ul S_i\}$ is a \v Cech 
$0$--cochain of the sheaf of $\SO(3)$ valued functions matching as
$$
\ul S_\alpha=\ul R_{\alpha\beta}\ul S_\beta
\eqno(6.2)
$$
under changes of trivialization of $E$, $\{\theta_i\}$ is an $\Omega^1$ 
\v Cech $0$--cochain, $\{\omega_{ij}\}$ is an $\Omega^0$ \v Cech $1$--cochain, 
both independent from the trivialization of $E$ used, and $\lambda_{\ul S i}$ 
is defined as in (2.16) with $\ul U$ replaced by $\ul S_i$.
More generally, $\scri C$ is called diagonalizable if it is 
equivalent to a Cho structure diagonalizable in the restricted 
sense just defined.

\rema{6.2}
It is easy to see that the restrictions imposed on $\ul S_i$,
$\theta_i$, $\omega_{ij}$ ensures that (3.1)--(3.4) are satisfied.
We note that (6.1b), (6.1c) are implied by (6.1a), as is easy to see from 
(2.16), (3.3), (4.2).

The connection $\ul A_i$ above has the space fixed form of ref. \ref{21}.

\prop{6.3} A diagonalizable Cho structure $\scri C$ is almost fine.

\proof
By rem. 5.9, one can assume that $\scri C$ satisfies (6.1)
without loss of generality.
Using (4.3), (4.6), (4.10), (6.1), one verifies directly that the 
associated length $3$ Deligne $3$--cocycle 
$(\{\eta_i\}$, $\{\psi_{ij}\},\{\phi_{ijk}\},\{K_{ijkl}\})$ 
is given by 
$$
\eqalignno{\vphantom{1\over 2}
\eta_i&=d\theta_i,&(6.3a)\cr
\vphantom{1\over 2}
\psi_{ij}&=\delta\theta_{ij}+d\omega_{ij},&(6.3b)\cr
\vphantom{1\over 2}
\phi_{ijk}&=\delta\omega_{ijk}+ l_{ijk},&(6.3c)\cr
\vphantom{1\over 2}
K_{ijkl}&=\delta l_{ijkl},&(6.3d)\cr
}
$$
where $\{l_{ijk}\}$ is some $2\pi\Bbb Z$ \v Cech $2$--cochain.
From (A.5),  it follows that 
$(\{\eta_i\},\{\psi_{ij}\},\{\phi_{ijk}\}$, $\{K_{ijkl}\})$
is a Deligne $3$--coboundary. Hence, the 
Deligne class $D_{\scri C}$ vanishes. By prop. 5.10, 
eq. (5.12), one concludes that $\scri C$ is almost fine. 
\hfill $\square$

\titlecps{The diagonalization theorem}

The natural question arises about under which conditions a Cho
structure $\scri C$ is diagonalizable. The answer is that it always is.

\theo{6.4}
Every Cho structure $\scri C$ is diagonalizable, thus almost fine, thus
equivalent to a fine Cho structure. In particular, for every
Cho structure $\scri C$, the Deligne class $D_{\scri C}$ is trivial
$$
D_{\scri C}=0
.\eqno(6.4)
$$

\proof 
To understand the essence of the
matter, let us study the following closely related problem. Given an $\Bbb E_3$
valued field $\ul n$ defined on an open set $O$ of $\Bbb R^p$ 
such that $\ul n^2=1$ and a vector $\ul n_0\in \Bbb E_3$ with 
$\ul n_0{}^2=1$, find an $\SO(3)$ valued function $\ul R$ on $O$ such that 
$$
\ul n=\ul R\ul n_0
.\eqno(6.5)
$$
It is not difficult to find the general expression of such an $\ul R$:
$$
\ul R=\ul R_{\ul n\leftarrow \ul n_0}\exp(-\varphi(\ul n)\star\ul n_0)
,\eqno(6.6)
$$
where
$$
\ul R_{\ul n\leftarrow \ul n_0}
=\ul 1+{1\over 1+\ul n\cdot\ul n_0}[\star(\ul n\times \ul n_0)]^2
+\star(\ul n\times \ul n_0)
.\eqno(6.7)
$$
and $\varphi(\ul n)$ is an arbitrary function possibly depending on $\ul n$.
The point is that $\ul R_{\ul n\leftarrow \ul n_0}$ is singular 
(though bounded) at those point $x\in O$ where $\ul n(x)=-\ul n_0$. 
These singularity cannot be compensated by a judicious choice of 
$\varphi(\ul n)$, as is easy to see from (4.18). So, when the field 
$\ul n$ takes all possible values in the unit sphere $S^2(\Bbb E_3)$ 
of $\Bbb E_3$, a regular $\SO(3)$ valued function $\ul R$ on $O$ 
fulfilling (6.4) cannot exist. However, it is easy to see that
every point $x\in O$ has an open neighborhood $O_x\subseteq O$ 
such that the range of $\ul n|_{O_x}$ is a proper subset of 
$S^2(\Bbb E_3)$. Indeed, if there were a point $x\in O$ such that for 
all open neighborhoods $O_x\subseteq O$ the range of $\ul n|_{O_x}$ were
the whole $S^2(\Bbb E_3)$, $\ul n$ would be singular at $x$, 
while $\ul n$ is regular on $O$. We conclude that $O$ has an 
open cover $\{O_a\}$, such that the problem posed has solution on each 
$O_a$ separately. 

From the discussion of the previous paragraph, it follows that 
for every Cho structure  ${\scri C}=(\{\ul n_i\},\{\ul A_i\},
\{\ul T_{ij}\})$ of $E$ subordinated to the open cover $\{O_i\}$, there is a
refinement $\{\bar O_{\bar\imath}\}$ of $\{O_i\}$ and a refinement map  
$f$ such that the associated refinement $\bar {\scri C}=(\{\ul {\bar n}_
{\bar\imath}\},\{\ul {\bar A}_{\bar\imath}\},
\{\ul {\bar T}_{\bar\imath\bar\jmath}\})$ satisfies (6.1) (cfr. def. 3.3). 
\hfill $\square$

Next we, explore the implications of this important result.

\titlecps{Examples}

\exa {6.5} {\it A small monopole loop in a single instanton background}

This example was illustrated in sect. 3. The Cho structure $\scri C$ defined 
in (3.15)--(3.17) is not manifestly diagonalizable. But it actually is
by theor. 6.4.

\exa {6.6} {\it A Higgs field on a $4$--torus}

Also this example was illustrated in sect. 3. 
The Cho structure $\scri C$ defined (3.22)--(3.24) is evidently 
diagonalizable.


\titlebf {7. Monopole and instanton charge, twist sectors and vortices}

We want now find expressions for the monopole and instanton charge
$m_{\scri C}$, $i_{\scri C}$ associated with a fine Cho structure 
${\scri C}$ and analyze their properties and relations.

\titlecps{Monopole and instanton charge}

Let $M$, $P$ and $E$ be as in sect. 3.
Let ${\scri C}$ be a fine Cho structure of $E$ (cfr. def. 5.1). 
Let $f_{\scri C}\in\Omega^2_{c2\pi\Bbb Z}(M)$ be the Abelian curvature 
of $\scri C$ (cfr. def. 5.6).

\defn{7.1} Let $\Sigma\in Z^s_2(M)$ be a finite singular $2$--cycle of $M$ \ref{33}.
\footnote{}{}\footnote{${}^3$}{
Here and in the following, we denote by $C^s_p(M)$, $Z^s_p(M)$, $B^s_p(M)$
the groups of $p$--dimensional singular chains, cycles and boundaries of $M$, 
respectively, and by $H^s_p(M)$ the degree $p$ singular homology of $M$.
Further, we denote by $C^s_p$ the precosheaf of $p$--dimensional singular 
chains \ref{36}.} The {\it monopole charge} $m_{\scri C}(\Sigma)$
of ${\scri C}$ in $\Sigma$ is
$$
m_{\scri C}(\Sigma)={1\over 4\pi}\int_\Sigma f_{\scri C}.
\eqno(7.1)
$$
Similarly, let $\Omega\in Z^s_4(M)$ be a finite singular 4--cycle of $M$.
The {\it instanton charge } $i_{\scri C}(\Omega)$ of ${\scri C}$ in $\Omega$ is
$$
i_{\scri C}(\Omega)={1\over 16\pi^2}\int_\Omega f_{\scri C}{}^2.
\eqno(7.2)
$$

It is important to realize that the objects just defined are indeed suitable 
generalizations of the customary physical objects carrying the same names. 
Indeed, 
$$
{1\over 4\pi}f_{\scri C}
=-{1\over 4\pi}\ul F_{\ul A_{\scri C}}\cdot\ul n_{\scri C}
={1\over 2\pi}\tr\big(F_{A_{\scri C}}n_{\scri C}\big)
,\eqno(7.3)
$$
$$
{1\over 16\pi^2}f_{\scri C}{}^2
={1\over 16\pi^2}\ul F_{\ul A_{\scri C}}\cdot\ul F_{\ul A_{\scri C}}
=-{1\over 8\pi^2}\tr\big(F_{A_{\scri C}}F_{A_{\scri C}}\big)
,\eqno(7.4)
$$
by prop. 5.5, where the trace is over the fundamental representation of 
$\goth s\goth u(2)$ (cfr. sect. 2). 
This justifies the identification of $m_{ {\scri C}}$,
$i_{\scri C}$ as monopole and instanton charge, respectively.

\proof From prop. 5.5, eq. (5.5), one has indeed that 
$\ul F_{\ul A_{\scri C}}=- f_{\scri C}\,\ul n_{\scri C}$. 
\hfill $\square$

In the physical applications where $M$ is a compact oriented $4$--dimensional
manifold, the $4$--cycle $\Omega$ is a representative of the fundamental class 
of $M$. However, the following treatment applies to more general situations.

\rema {7.2}
Since $f_{\scri C}\in\Omega^2_{c2\pi\Bbb Z}(M)$, 
$m_{\scri C}(\Sigma)\in\Bbb Z/2$, $i_{\scri C}(\Omega)\in\Bbb Z/4$. However, if 
$f_{\scri C}/2\in\Omega^2_{c2\pi\Bbb Z}(M)$, then $m_{\scri C}(\Sigma)$, 
$i_{\scri C}(\Omega)\in\Bbb Z$. 
 
\rema{7.3}
By standard arguments \ref{33}, $m_{\scri C}$ induces a homomorphism 
$m_{\scri C}:H^s_2(M) \rightarrow\Bbb Z/2$ on the degree 2 singular homology of $M$. 
Similarly, $i_{\scri C}$ induces a homomorphism $i_{\scri C}:H^s_4(M)
\rightarrow\Bbb Z/4$ on the degree 4 singular homology of $M$. To emphasize this, 
one may write $m_{\scri C}([\Sigma])$, $i_{\scri C}([\Omega])$
in (7.1), (7.2).

\rema{7.4}
Since $f_{\scri C}$ depends on ${\scri C}$ only through its 
fine equivalence class $\langle{\scri C}\rangle$, by prop. 5.6, so do
$m_{\scri C}$, $i_{\scri C}$. 
For this reason, one occasionally writes $m_{\langle{\scri C}\rangle}$, 
$i_{\langle{\scri C}\rangle}$ to emphasize this fact.

\titlecps{Discussion}

Suppose that the Cho structure 
${\scri C}=(\{\ul n_i\},\{\ul A_i\},\{\ul T_{ij}\})$ is subordinated 
to the open cover $\{O_i\}$. 
From (4.4) and the relation $\ul F_{\ul A i}=-f_{\scri C}|_{O_i}\ul n_i$, 
one has 
$$
f_{\scri C}|_{O_i}
=\hbox{$1\over 2$}\ul n_i\cdot d\ul n_i\times d\ul n_i
-d(\ul A_0\cdot\ul n_i)+da_i.
\eqno(7.5)
$$
We note that $({1\over 2}\ul n_i\cdot d\ul n_i\times d\ul n_i)^2=0$ 
identically, as is easy to see. From this fact and (7.5)
$$
f_{\scri C}{}^2|_{O_i}
=\big(\ul n_i\cdot d\ul n_i\times d\ul n_i
-d(\ul A_0\cdot\ul n_i)+da_i\big)\big(-d(\ul A_0\cdot\ul n_i)+da_i\big).
\eqno(7.6)
$$
This formula shows the importance of the terms  $-d(\ul A_0\cdot\ul n_i)+da_i$
in (7.5) to yield a non vanishing instanton number. 
In the formulation of ref. \ref{20}, terms like these are 
associated with Dirac strings, sheets etc. and are distributional.
In our formulation. Dirac strings, sheets etc. are traded for cocycles 
specifying glueing of locally defined fields. 
In turn, glueing requires these terms. 

From (7.5), we see that 
$w_{\scri C}=m_{\scri C}$ is a generalization of the 
customary winding number. Similarly, one can relate the instanton charge 
$i_{\scri C}$ to a suitable generalization of the Hopf 
invariant. To this end, we assume that the open sets of the cover 
$\{O_i\}$ are contractible. 
As $f_{\scri C}\in\Omega^2_c(M)$ and $O_i$ is contractible, one has
$$
f_{\scri C}|_{O_i}=dv_i
,\eqno(7.7)
$$
for some $v_i\in\Omega^1(O_i)$, by Poincar\'e's lemma. 
Since $f_{\scri C}/4\pi$ is a winding number density,
as explained above, $v_idv_i/16\pi^2$ is a local Hopf invariant density
on $O_i$. Now, we note that
$$
f_{\scri C}{}^2|_{O_i}=d(v_idv_i)
.\eqno(7.8)
$$
Let $\Omega\in Z^s_4(M)$ be an $\{O_i\}$--small finite singular 4--cycle of 
$M$ \ref{33}. Then, there are a $C^s_4$ \v Cech $0$--chain $\{U_i\}$ and 
a $C^s_3$ \v Cech $1$--chain $\{V_{ij}\}$ such that 
$$
\Omega=\sum_iU_i, \quad bU_i=\sum_jV_{ji}
,\eqno(7.9)
$$
where $b$ is the singular boundary operator \ref{33}. 
Then, using (7.8), (7.9) and Stokes' theorem, one easily shows that
$$
i_{\scri C}(\Omega)={1\over 32\pi^2}\sum_{ij}
\int_{V_{ij}}\big(v_jdv_j-v_idv_i\big)
.\eqno(7.10)
$$
Intuitively, $i_{\scri C}(\Omega)$ is given by the net discontinuity of
the local Hopf invariant densities $v_idv_i/16\pi^2$ across the $V_{ij}$.
This is a cohomological interpretation of the calculations of ref. \ref{37}.
Exploiting barycentric subdivision, the above construction can be easily 
generalized to finite singular 4--cycles $\Omega\in Z^s_4(M)$ which are not
necessarily $\{O_i\}$--small \ref{33}.

\titlecps{Liftability of $\scri C$}

Let $\scri C$ be a fine Cho structure. The $\SO(3)$ vector bundle 
$E_{\scri C}$ associated with $\scri C$ (cfr. prop. 5.5) 
is not liftable to an $\SU(2)$ one in general. Thus, we expect
the second Stiefel--Whitney class $w_2(E_{\scri C})\in H^2(M,\Bbb Z_2)$ 
of $E_{\scri C}$ to play a role. 

\defprop{7.5} The {\it Stiefel--Whitney class} 
$\varepsilon_{\scri C}\in H^2(M,\Bbb Z_2)$ of $\scri C$ is
$$
\varepsilon_{\scri C}=w_2(E_{\scri C}).
\eqno(7.11)
$$
$\varepsilon_{\scri C}$ depends on $\scri C$ only through its fine equivalence 
class $\langle{\scri C}\rangle$.

For this reason, one may write $\varepsilon_{\langle{\scri C}\rangle}$ 
to stress this fact.

\proof The statement follows trivially from the last part of prop. 5.5.
\hfill $\square$

\defprop{7.6} The $\SO(3)$ bundle $E_{\scri C}$ associated with $\scri C$ 
is liftable to an $\SU(2)$ one precisely when $\varepsilon_{\scri C}$ vanishes. 
When this happens, the fine Cho structure $\scri C$ is called {\it liftable}.

\proof The proof is trivial \ref{34} \hfill $\square$

\rema {7.7} If $\scri C$ is liftable, there are in general several 
$\SU(2)$ lifts of $E_{\scri C}$. As is  well known, these 
are classified cohomologically by $H^1(M,\Bbb Z_2)$ \ref{34}. 

\prop{7.8} One has 
$$
\varepsilon_{\scri C}=c(e_{\scri C})\qquad \hbox{mod $2$}
\eqno(7.12)
$$
where $c(l)$ denotes the 1st Chern class of a line bundle $l$
and $e_{\scri C}$ is defined in prop. 5.5. Thus,  
$\scri C$ is liftable if and only if $e_{\scri C}$ is a square. In that case,
the monopole charge $m_{\scri C}$ and the instanton charge $i_{\scri C}$ 
are $\Bbb Z$ valued.

\proof According to prop. 5.5, the $\SO(3)$ bundle $E_{\scri C}$ decomposes according 
to (5.6), (5.7). It is known that $w_2(E_{\scri C})=c(e_{\scri C})$ mod $2$
(see ref. \ref{34}). So (7.12) is clear. Under the natural homomorphism 
$H^2(M,\Bbb Z)\rightarrow H^2(M,\Bbb R)$, $c(e_{\scri C})$ is represented by
$[f_{\scri C}/2\pi]$. So, when $\varepsilon_{\scri C}$ vanishes, $f_{\scri C}/2
\in \Omega^2_{c2\pi\Bbb Z}(M)$. Thus, by rem 7.2, $m_{\scri C}$, $i_{\scri C}$ 
are integer valued. \hfill $\square$

Our framework naturally accommodates half integral monopole charges
a quarter integral instanton charges.
The reason why $m_{\scri C}$, $i_{\scri C}$ are respectively 
$\Bbb Z/2$ and $\Bbb Z/4$ valued can be traced to the fact that 
$\scri C$ is generally non liftable.

\rema{7.9} View $\scri C$ as an almost fine Cho structure (cfr. def. 5.8). 
Then, the associated set of fine equivalence classes of fine Cho structures 
${\sans F}_{\langle{\scri C}\rangle}$ (cfr. def. 5.12) is 
partitioned in sectors depending on the value of 
$\varepsilon_{\langle{\scri C}'\rangle}\in H^2(M,\Bbb Z_2)$. 

These sectors answer to the twisted sectors of refs.
\ref{16--18}. This is particularly clear from the analysis of sect. 2. 
Thus, they are also related to center vortices, as explained in
the introduction.

\titlecps{Other topological features}

Let $\scri C$, ${\scri C}'$ be equivalent fine Cho structures. 
If their Abelian curvatures $f_{\scri C}$, $f_{{\scri C}'}$ are equal, 
one cannot conclude that ${\scri C}$, ${\scri C}'$ are finely equivalent 
in general.

\defn{7.10} Let $\scri C$ be an almost fine Cho structure. We denote by 
${\sans F}_{{\scri C}f}$ the subset of ${\sans F}_{\scri C}$ of fine equivalence 
classes of fine Cho structures equivalent to $\scri C$ with assigned 
Abelian curvature $f\in\Omega^2_{c2\pi\Bbb Z}(M)$. 

\prop{7.11} Let $\scri C$ be an almost fine Cho structure and
$f\in\Omega^2_{c2\pi\Bbb Z}(M)$. Then, there exists
a bijection ${\sans F}_{{\scri C}f}\cong H^1(M,\Bbb T)$

\proof There is an exact sequence 
$$
\matrix{
& & & {}_\varrho & & {}_\varsigma & & & \cr
0&\rightarrow &H^1(M,\Bbb T)&
\rightarrow &H^2(M, D(2)^\bullet) &\rightarrow & \Omega^2_{c2\pi\Bbb Z}(M)&
\rightarrow &0,\cr
}
\eqno(7.13)
$$
where the map $\varsigma$ associates to any class in $H^2(M,D(2)^\bullet)$
its curvature $\Omega^2_{c2\pi\Bbb Z}(M)$ \ref{32}. 

Without loss of generality we can assume that $\scri C$ is fine. 
Then, we can identify ${\sans F}_{\scri C}$ 
with $H^2(M,D(2)^\bullet)$ using the bijection ${\sans j}_{\scri C}{}^{-1}$
defined in the proof of theor. 5.13. It is easy to see that 
$$
\varsigma\circ {\sans j}_{\scri C}{}^{-1}(\langle {\scri C}'\rangle)
=f_{{\scri C}'}-f_{\scri C}
\eqno(7.14)
$$
for a fine class $\langle {\scri C}'\rangle\in {\sans F}_{\scri C}$.
Hence, ${\sans F}_{{\scri C}f}$ is identified via ${\sans j}_{\scri C}{}^{-1}$
with $\ker\varsigma$. By the exactness of (7.13), 
$\ker\varsigma\cong\ran\varrho\cong H^1(M,\Bbb T)$.
The statement follows. \hfill $\square$

\rema{7.12}
$H^1(M,\Bbb T)$ has a well known interpretation. It is the group of flat 
$\Bbb T$ principal bundle on $M$. $H^1(M,\Bbb T)$ is aptly described by the exact
sequence
$$
\matrix{& & & {}_\chi& & {}_c & & & \cr
0&\rightarrow & H_{dR}^1(M)/H_{dR\Bbb Z}^1(M) &
\rightarrow & H^1(M,\Bbb T)& \rightarrow &\Tor H^2(M,\Bbb Z)&
\rightarrow & 0.\cr
}
\eqno(7.15)
$$
Here, $H_{dR}^1(M)$ is the degree $1$ de Rham cohomology. $H_{dR\Bbb Z}^1(M)$
is the integral lattice in $H_{dR}^1(M)$. $\chi$ is essentially the map
$\exp(2\pi\sqrt{-1}~\cdot~)$ in the \v Cech formulation of cohomology.
$c$ is the Chern class homomorphism. The degree $2$ torsion 
$\Tor H^2(M,\Bbb Z)$ is the kernel of the 
natural homomorphism $H^2(M,\Bbb Z)\rightarrow H^2(M,\Bbb R)$.

\proof (7.14) can be easily deduced from the long exact sequence of cohomology 
associated with the standard short exact sequence of sheaves 
$0\rightarrow \Bbb Z \rightarrow \Bbb R \rightarrow \Bbb T \rightarrow 0$. 
\hfill $\square$

The above discussion shows that the monopole and instanton charges 
$m_{\scri C}$, $i_{\scri C}$ do not completely characterize the monopole 
configuration associated with a fine Cho structure $\scri C$. 
There are further topological features associated with the group 
$H^1(M,\Bbb T)$. In the simple case where $M$ is compact and the torsion 
$\Tor H^2(M,\Bbb Z)$ vanishes, $H^1(M,\Bbb T)$ is the torus ${\Bbb T}^{b_1}$,
where $b_1$ is the 1st Betti number of $M$.

\vskip.6cm
\par\noindent
{\bf Acknowledgments.} We are greatly indebted to R. Stora for useful discussion.

\vfill\eject

\titlebf {A. Smooth Deligne cohomology}

Our method relies heavily on the smooth version of Deligne cohomology.
Since Deligne cohomology is defined in terms of hypercohomology, we
review briefly below the basic facts about hypercohomology used in the 
paper. See ref. \ref{32,33} for background material.

Abstractly, the computation of the {\it hypercohomology} of a complex 
of sheaves of Abelian groups $K^\bullet$
$$
\matrix{&{}_{d_K}&&{}_{d_K}&&{}_{d_K}&\cr
K^0&\longrightarrow &K^1&\longrightarrow &\cdots &\longrightarrow & K^p
}
\eqno(A.1)
$$ 
requires the choice of an appropriate resolution $R^{\bullet\bullet}$ of 
$K^\bullet$
$$
\matrix{&\!\!\!\!{}_r&&{}_\delta&&{}_\delta&&{}_\delta&\cr
K^p&\!\!\!\!\longrightarrow &R^{0,p}&\longrightarrow &R^{1,p}&\longrightarrow 
&R^{2,p}&\longrightarrow& \cdots \cr
{~\atop d_K}{~\atop\big\uparrow}~~~~&&{~\atop d}
{~\atop\big\uparrow}~~~&&{~\atop d}{~\atop\big\uparrow}~~~&&
{~\atop d}{~\atop\big\uparrow}~~~&\cr
\vdots&&\!\!\vdots&&\!\!\vdots&&\!\!\vdots&\cr
{~\atop d_K}{~\atop\big\uparrow}~~~~&&{~\atop d}
{~\atop\big\uparrow}~~~&&{~\atop d}{~\atop\big\uparrow}~~~&&
{~\atop d}{~\atop\big\uparrow}~~~&\cr
&\!\!\!\!{}_r&&{}_\delta&&{}_\delta&&{}_\delta&\cr
K^1&\!\!\!\!\longrightarrow &R^{0,1}&\longrightarrow &R^{1,1}&\longrightarrow 
&R^{2,1}&\longrightarrow& \cdots \cr
{~\atop d_K}{~\atop\big\uparrow}~~~~&&{~\atop d}
{~\atop\big\uparrow}~~~&&{~\atop d}{~\atop\big\uparrow}~~~&&
{~\atop d}{~\atop\big\uparrow}~~~&\cr
&\!\!\!\!{}_r&&{}_\delta&&{}_\delta&&{}_\delta&\cr
K^0&\!\!\!\!\longrightarrow &R^{0,0}&\longrightarrow &R^{1,0}&\longrightarrow 
&R^{2,0}&\longrightarrow&\cdots \cr
}
\eqno(A.2)
$$ 
e. g. an injective resolution. $R^{\bullet\bullet}$ is a double complex
of sheaves of Abelian groups. 
The hypercohomology of $K^\bullet$, $H^\bullet(M,K^\bullet)$, is 
the cohomology of the complex of Abelian groups ${\rm Tot}R^\bullet(M)$
$$
\matrix{&{}_{d_{{\rm Tot}RM}}&&{}_{d_{{\rm Tot}RM}}&&{}_{d_{{\rm Tot}RM}}&\cr
{\rm Tot}R^0(M)&\longrightarrow &{\rm Tot}R^1(M)&\longrightarrow &\cdots &
\longrightarrow & {\rm Tot}R^p(M),
}
\eqno(A.3)
$$ 
where ${\rm Tot}R^k=\bigoplus_{l=0}^{\min(p,k)}R^{k-l,l}$
and $d_{{\rm Tot}R}=\delta+(-1)^\partial d$ with $\partial$ 
denoting the horizontal degree.
$H^\bullet(M,K^\bullet)$ is independent from the choice of the resolution 
up to isomorphism. 

From the above definition, it follows that a hypercohomology class
$c$ of $H^k(M,K^\bullet)$ is represented by a $k$--cocycle $\gamma$ 
of ${\rm Tot}R^k(M)$ defined up to a $k$--coboundary. 
A $k$--cocycle $\gamma$ of ${\rm Tot}R^k(M)$ is a sequence 
$(\gamma^{k-l,l})_{l=0,1,\ldots,\min(p,k)}$, 
where $\gamma^{k-l,l}\in R^{k-l,l}(M)$ satisfy
$$
\delta\gamma^{k-l,l}=d\gamma^{k-l+1,l-1}, \quad l=0,1,\ldots,\min(p,k)+1,
\eqno(A.4)
$$
with $\gamma^{k+1,-1}=0$, $\gamma^{k-\min(p,k)-1,\min(p,k)+1}=0$.
A $k$--coboundary $\beta$ of ${\rm Tot}R^k(M)$ is a  
$k$--cocycle of ${\rm Tot}R^k(M)$ of the form 
$$
\beta^{k-l,l}=\delta c^{k-1-l,l}+dc^{k-l,l-1}, 
\quad l=0,1,\ldots,\min(p,k),
\eqno(A.5)
$$
where $c^{k-1-l,l}\in R^{k-1-l,l}(M)$, $l=0,1,\ldots,\min(p,k-1)$, 
and $c^{k,-1}=0$, $c^{-1,k}=0$ ($k\leq p$).

In practical calculations, it is convenient to use a \v Cech resolution 
of $K^\bullet$, which is constructed as follows. Let $\{O_i\}$
be an open cover of $M$. We set $R^{n,l}$ is the sheaf of \v Cech 
$n$--cochains of the sheaf $K^l$. 
For an open set $U$, an element $\gamma^{n,l}\in R^{n,l}(U)$ is a 
collection $\{\gamma^{n,l}{}_{i_0\cdots i_n}\}$, 
where $\gamma^{n,l}{}_{i_0\cdots i_n}\in K^l(O_{i_0\ldots i_n}\cap U)$ 
and $\gamma^{n,l}{}_{i_0\cdots i_n}$ is totally antisymmetric in the cover 
indices $i_0$, ..., $i_n$. 
\footnote{}{}\footnote{${}^41$}{Below, we denote by $O_{ij}$, $O_{ijk}$, ... 
the non empty intersections $O_i\cap O_j$, $O_i\cap O_j\cap O_k$, ...,
respectively. }
The inclusion $r$ and the coboundary operators 
$\delta$ and $d$ are defined as 
$$
(r\alpha^l)_i=\alpha^l|_{O_i\cap U},
\eqno(A.6)
$$
$$
(\delta\gamma^{n,l})_{i_0\cdots i_{n+1}}
=\sum_{r=0}^{n+1}(-1)^r\gamma^{n,l}{}_{i_0\cdots \not i_r\cdots i_{n+1}}
|_{O_{i_0\ldots i_{n+1}}\cap U},
\eqno(A.7)
$$
$$
(d\gamma^{n,l})_{i_0\cdots i_n}=d_K\gamma^{n,l}_{i_0\cdots i_n}.
\eqno(A.8)
$$
for $\alpha^l\in K^l(U)$ and $\gamma^{n,l}\in R^{n,l}(U)$.
For a generic cover $\{O_i\}$, the cohomology of the complex
${\rm Tot}R^\bullet(M)$ depends on $\{O_i\}$ and does not directly
compute the hypercohomology $H^\bullet(M,K^\bullet)$. To that end, 
it is necessary to perform the direct limit of the cohomology of 
${\rm Tot}R^\bullet(M)$ with respect to refinements of $\{O_i\}$. 
In favorable conditions, there is a class of covers, called good
covers, such that, when the cover $\{O_i\}$ is good, the cohomology 
of ${\rm Tot}R^\bullet(M)$ is isomorphic to the hypercohomology 
$H^\bullet(M,K^\bullet)$ and there is no need of the direct limit.

Let $p\in\Bbb N$. The {\it length $p$ smooth Deligne complex} $D(p)^\bullet$ 
is the complex of sheaves
$$
\matrix{&{}_j&&{}_d&&{}_d&&{}_d&\cr
2\pi\Bbb Z&\longrightarrow &\Omega^0&\longrightarrow 
&\Omega^1&\longrightarrow &\cdots &\longrightarrow&\Omega^{p-1}.\cr
}
\eqno(A.9)
$$
Here, $\Bbb Z$ is the sheaf of locally constant integer valued functions
on $M$. $\Omega^k$ is the sheaf of real valued $k$--forms of $M$. 
$j$ is the natural injection and $d$ is the customary de Rham differential. 
$2\pi\Bbb Z$ is put in degree $0$. 
\footnote{}{}\footnote{${}^52$}{Our definition of the smooth Deligne complex 
differs from that of ref. \ref{32}, where one puts the 
sheaf $\Bbb Z(p)=(2\pi \imu)^p\Bbb Z$ in degree 0 and the sheaf
$\Omega^{k-1}_{\Bbb C}$ of complex valued $k-1$--forms in degree $k>0$.}
The {\it length $p$ smooth Deligne cohomology} is by definition the 
hypercohomology of the smooth Deligne complex $D(p)^\bullet$, 
$H^\bullet(M,D(p)^\bullet)$. 
Below, we shall deal with Deligne cohomology using the \v Cech resolution 
$R^{\bullet\bullet}$ of $D(p)^\bullet$ associated with an open cover 
$\{O_i\}$, as outlined above. 
When necessary, we shall assume that $\{O_i\}$ is good. In the case considered
here, an open cover $\{O_i\}$ is good if all the non empty finite 
intersections $O_{i_0\cdots i_k}$ are contractible. A cohomology class $c$ of 
$H^k(M,D(p)^\bullet)$ is then represented by a $k$--cocycle $\gamma$ 
of ${\rm Tot}R^k(M)$ defined up to a $k$--coboundary (cfr. eqs. (A.4)--(A.5)). 
In the following, we shall call such cocycles, {\it length $p$ 
Deligne $k$--cocycles}.

Some of the Deligne cohomology groups classify certain differential 
topological structures having the circle group $\Bbb T$ as structure group
\ref{32}. Below, we describe the ones which are relevant in the paper.

$H^2(M,D(2)^\bullet)$ is isomorphic to the group of isomorphism 
classes of Hermitian line bundles with Hermitian connection. 
Indeed, a class of $H^2(M,D(2)^\bullet)$ is represented by 
a $2$--cocycle $(\{a_i\},\{f_{ij}\},\{m_{ijk}\})$, 
where $\{a_i\}$ is an $\Omega^1$ \v Cech $0$--cochain, 
$\{f_{ij}\}$ is an $\Omega^0$ \v Cech $1$--cochain and $\{m_{ijk}\}$ is 
a $2\pi\Bbb Z$ \v Cech $2$--cochain, satisfying the relations
$$
\eqalignno{\vphantom{1\over 2}
\delta a_{ij}&=df_{ij},&(A.10a)\cr
\vphantom{1\over 2}
\delta f_{ijk}&=m_{ijk},&(A.10b)\cr
\vphantom{1\over 2}
\delta m_{ijkl}&=0.&(A.10c)\cr
}
$$
These relations imply that $\{\exp(\imu f_{ij})\}$ is a $\ul{\Bbb T}$
\v Cech $1$--cocycle representing a Hermitian line bundle $L$ and 
$\{\imu a_i\}$ is a $\imu \Omega^1$ \v Cech $0$--cochain corresponding to
a Hermitian connection $A$ of $L$. 
\footnote{}{}\footnote{${}^6$}{Below, for any Lie group $G$, 
$\ul G$ denotes the sheaf of $G$ valued functions.}
If we replace the $2$--cocycle 
$(\{a_i\},\{f_{ij}\},\{m_{ijk}\})$ by a cohomologous one, 
we obtain an equivalent Hermitian line bundle with Hermitian 
connection. The $\Bbb Z$ \v Cech $2$--cocycle $\{m_{ijk}/2\pi\}$ represents 
the Chern class $c(L)\in H^2(M,\Bbb Z)$ of $L$. The closed $2$--form 
with integer periods $F\in\Omega_{c\Bbb Z}^2(M)$,
defined by $F|_{O_i}=da_i/2\pi$, is the curvature of $A$.
$c(L)$ and $F$ depend only on the isomorphism class of $L$, $A$,
hence on the corresponding Deligne cohomology class.  

Similarly, $H^3(M,D(3)^\bullet)$ is isomorphic to the group 
of isomorphism classes of Hermitian gerbes with Hermitian connective 
structure and curving.
Indeed, a class of $H^3(M$, $D(3)^\bullet)$ is represented by 
a $3$--cocycle $(\{h_i\},\{g_{ij}\},
\{f_{ijk}\},\{m_{ijkl}\})$, where $\{h_i\}$ is an $\Omega^2$ \v Cech 
$0$--cochain, $\{g_{ij}\}$ is an $\Omega^1$ \v Cech $1$--cochain, 
$\{f_{ijk}\}$ is an $\Omega^0$ \v Cech $2$--cochain and $\{m_{ijkl}\}$ 
is a $2\pi\Bbb Z$ \v Cech $3$--cochain satisfying the relations
$$
\eqalignno{\vphantom{1\over 2}
\delta h_{ij}&=dg_{ij},&(A.11a)\cr
\vphantom{1\over 2}
\delta g_{ijk}&=df_{ijk},&(A.11b)\cr
\vphantom{1\over 2}
\delta f_{ijkl}&=m_{ijkl},&(A.11c)\cr
\vphantom{1\over 2}
\delta m_{ijklm}&=0.&(A.11d)\cr
}
$$
These relations imply that $\{\exp(\imu f_{ijk})\}$ is a $\ul{\Bbb T}$
\v Cech $2$--cocycle representing a Hermitian gerbe $H$ and  
$\{\imu g_{ij}\}$, $\{\imu h_i\}$ are a $\imu \Omega^1$ \v Cech $1$--cochain 
and a $\imu \Omega^2$ \v Cech $0$--cochain corresponding to
a Hermitian connective structure and curving $C$ of $H$,
respectively. If we replace the $3$--cocycle 
$(\{h_i\},\{g_{ij}\},\{f_{ijk}\},\{m_{ijkl}\})$ by a cohomologous one, 
we obtain an equivalent Hermitian gerbe with Hermitian 
connective structure and curving. The $\Bbb Z$ \v Cech $3$--cocycle 
$\{m_{ijkl}/2\pi\}$ represents the Dixmier--Douady 
class $d(H)\in H^3(M,\Bbb Z)$ of $H$. The closed $3$--form 
with integer periods $G\in\Omega_{c\Bbb Z}^3(M)$,
defined by $G|_{O_i}=dh_i/2\pi$, is the curvature of $C$.
$d(H)$ and $G$ depend only on the isomorphism class of $H$, $C$,
hence on the corresponding Deligne cohomology class.

\vfill\eject

\vskip.6cm
\titlebf {REFERENCES}
\vskip.6cm

\item{[1]}
Y.~Nambu,
``Strings, Monopoles and Gauge Fields'',
Phys.\ Rev.\ D {\bf 10} (1974) 4262.

\item{[2]}
G.~Parisi,
``Quark Imprisonment and Vacuum Repulsion'',
Phys.\ Rev.\ D {\bf 11} (1975) 970.

\item{[3]}
S.~Mandelstam,
``Vortices and Quark Confinement in Nonabelian Gauge Theories,''
Phys.\ Rept.\ {\bf 23} (1976) 245.

\item{[4]}
A.~M.~Polyakov,
``Quark Confinement and Topology of Gauge Groups'',
Nucl.\ Phys.\ B {\bf 120} (1977) 429.

\item{[5]}
G.~'t Hooft,
``A Property of Electric and Magnetic Flux in Nonabelian Gauge Theories'',
Nucl.\ Phys.\ B {\bf 153} (1979) 141.

\item{[6]}
G.~'t Hooft,
``Topology of the Gauge Condition and New Confinement Phases in Nonabelian 
Gauge Theories'',
Nucl.\ Phys.\ B {\bf 190} (1981) 455.

\item{[7]}
G.~'t Hooft,
``On the Phase Transition Towards Permanent Quark Confinement'',
Nucl.\ Phys.\ B {\bf 138} (1978) 1.

\item{[8]}
Y.~Aharonov, A.~Casher and S.~Yankielowicz,
``Instantons and Confinement'',
Nucl.\ Phys.\ B {\bf 146} (1978) 256.

\item{[9]}
J.~M.~Cornwall,
``Quark Confinement and Vortices in Massive Gauge Invariant QCD'',
Nucl.\ Phys.\ B {\bf 157} (1979) 392.

\item{[10]}
H.~B.~Nielsen and P.~Olesen,
``A Quantum Liquid Model for the QCD Vacuum: Gauge and Rotational Invariance 
of Domained and Quantized Homogeneous Color Fields'',
Nucl.\ Phys.\ B {\bf 160} (1979) 380.

\item{[11]}
G.~Mack and V.~B.~Petkova,
``Comparison of Lattice Gauge Theories with Gauge Groups Z(2) and SU(2)'',
Annals Phys.\  {\bf 123} (1979) 442.

\item{[12]}
L.~Del Debbio, M.~Faber, J.~Greensite and S.~Olejnik,
``Center Dominance and Z(2) Vortices in SU(2) Lattice Gauge Theory,''
Phys.\ Rev.\ D {\bf 55} (1997) 2298, arXiv:hep-lat/9610005.

\item{[13]}
C.~Alexandrou, M.~D'Elia and P.~de Forcrand,
``The Relevance of Center Vortices'',
Nucl.\ Phys.\ Proc.\ Suppl.\  {\bf 83} (2000) 437,
arXiv:hep-lat/9907028.

\item{[14]}
P.~de Forcrand and M.~Pepe,
``Center Vortices and Monopoles without Lattice Gribov Copies'',
Nucl.\ Phys.\ B {\bf 598} (2001) 557, arXiv:hep-lat/0008016.

\item{[15]}
J.~Ambjorn, J.~Giedt and J.~Greensite,
``Vortex Structure vs. Monopole Dominance in Abelian Projected Gauge Theory'',
JHEP {\bf 0002} (2000) 033, arXiv:hep-lat/9907021.

\item{[16]}
L.~G.~Yaffe,
``Confinement in SU(N) Lattice Gauge Theories'',
Phys.\ Rev.\ D {\bf 21} (1980) 1574.

\item{[17]}
E.~T.~Tomboulis,
``The t Hooft Loop in SU(2) Lattice Gauge Theory'',
Phys.\ Rev.\ D {\bf 23} (1981) 2371.

\item{[18]}
P.~de Forcrand and O.~Jahn,
``Comparison of SO(3) and SU(2) Lattice Gauge Theory'',
Nucl.\ Phys.\ B {\bf 651} (2003) 125, arXiv:hep-lat/0211004.

\item{[19]}
H.~Reinhardt and T.~Tok,
``Abelian and center gauges in continuum Yang-Mills theory'',
arXiv:hep-th/0009205.

\item{[20]}
F.~Bruckmann,
``Instanton Induced Defects in Gauge Theory'',
Ph. D. thesis, University of Jena.

\item{[21]}
A.~Di Giacomo and M.~Mathur,
``Abelianization of SU(N) Gauge Theory with Gauge Invariant Dynamical  
Variables and Magnetic Monopoles'',
Nucl.\ Phys.\ B {\bf 531} (1998) 302, arXiv:hep-th/9802050.

\item{[22]}
H.~Reinhardt,
``Topology of Center Vortices'',
Nucl.\ Phys.\ B {\bf 628} (2002) 133, rXiv:hep-th/0112215.

\item{[23]}
Y.~M.~Cho,
``Extended Gauge Theory and its Mass Spectrum'',
Phys.\ Rev.\ D {\bf 23} (1981) 2415.

\item{[24]}
Y.~M.~Cho,
``A Restricted Gauge Theory'',
Phys.\ Rev.\ D {\bf 21} (1980) 1080.

\item{[25]}
T.~ T.~Wu and C.~N.~ Yang,
``Concept of non Integrable Phase Factors and Global Formulation of Gauge 
Fields''.
Phys.\ Rev.\ {\bf D12} (1975) 3845.

\item{[26]}
O.~Alvarez,
``Cohomology and Field Theory'', UCB-PTH-85/20
Plenary talk given at Symp. on Anomalies, Geometry and Topology, Argonne, 
IL, Mar 28-30, 1985.

\item{[27]}
O.~Alvarez,
``Topological Quantization and Cohomology'',
Commun.\ Math.\ Phys.\  {\bf 100} (1985) 279.

\item{[28]}
K.~Gawedzki,
``Topological Actions in Two-Dimensional Quantum Field Theories'',
in  Cargese 1987, proceedings, Nonperturbative Quantum Field Theory, 101.

\item{[29]}
P.~Deligne,
``Th\'eorie de Hodge'',
Inst. Hautes \'Etudes Sci. Publ. Math. {\bf 40} (1971) 5.

\item{[30]}
A.~A.~Beilinson,
``Higher regulators and values of L--functions'',
J. Soviet Math. {\bf 30} (1985) 2036.

\item{[31]}
A.~A.~Beilinson,
``Notes on Absolute Hodge Cohomology'',
in Contemp. Math. {\bf 55} Part I  Amer. Math. Soc. (1986).

\item{[32]}
J.~-L.~Brylinski,
``Loop Spaces, Characteristic Classes and Geometric Quantization'',
Birkh\"auser, 1993.

\item{[33]}
R.~Bott and L.~Tu,
``Differential Forms in Algebraic Topology'',
Springer Verlag, New York, 1982.

\item{[34]} S. Donaldson and P. Kronheimer, 
``The Geometry of Four-Manifolds'', 
Clarendon Press, Oxford 1990.

\item{[35]}
R.~C.~Brower, K.~N.~Orginos and C.~I.~Tan,
``Magnetic Monopole Loop for the Yang-Mills Instanton'',
Phys.\ Rev.\ D {\bf 55} (1997) 6313, arXiv:hep-th/9610101.

\item{[36]}
G.~E.~Bredon,
``Sheaf Theory'',
Springer Verlag, New York, 1991.

\item{[37]}
O.~Jahn,
``Instantons and Monopoles in General Abelian Gauges'',
J.\ Phys.\ A {\bf 33} (2000) 2997, arXiv:hep-th/9909004.

\bye


As already discussed, one is embedding 
the original $G$ gauge theory in a larger $G/Z$ gauge theory
comprising a number of sectors corresponding to the possible 
monodromy types of the gauge transformations $g$. The original $G$ gauge
theory corresponds to the trivial monodromy sector. 
The holonomy of a gauge field around any given 
loop is defined only up to elements of $Z$, depending on its twist sector, 
i. e. the gauge field has $G/Z$ rather $G$ holonomy for non trivial twist.
One can lift the holonomy of a gauge field of the former sector
from $G/Z$ to $G$, if one restricts oneself to loops which are 
not linked to the branching sheets of the multivalued gauge transformation. 
The branching sheets are vortex world sheets.  
the relevant gauge group is $G/Z$ rather than $G$.


After Abelian gauge fixing and Abelian projection, 
the gauge field $\ul A$ has the simple space fixed frame form 
$$
\ul A_{\rm sff}=a_{\rm sff}\ul n_0
,\eqno(1.1)
$$
where $\ul n_0$ is a fixed element of the Lie algebra $\goth s\goth u(2)$ 
viewed as a $3$--dimensional Euclidean vector of $\Bbb E_3$ normalized so that
$\ul n_0{}^2=1$ \ref{23}. 
The space fixed frame form of the gauge field is not the most 
interesting form in many respects. It may be convenient to undo the Abelian 
gauge fixing, but not the Abelian projection, by applying an arbitrary $\SU(2)$
gauge transformation to $\ul A_{\rm sff}$.


One can view an Abelian gauge as map that assigns to each gauge field $A$ a non 
vanishing Higgs field $\phi(A)$ transforming in the adjoint representation of 
the gauge group $G$, i. e. satisfying 
$$
\phi(A^g)=\Ad g^{-1}\phi(A), 
\eqno(1.1)
$$
for any gauge transformation $g$, where $A^g$ is the gauge transform of $A$. By 
definition, the gauge transformation $\bar g$ which carries $A$ into the Abelian 
gauge is the one which renders $\phi(A^{\bar g})$ $\goth h$ valued \ref{6}. 
$\bar g$ is defined up to right multiplication by an arbitrary $H$ valued guage 
transformation $h$, $\bar g\rightarrow \bar g h$. So, the Abelian gauge fixing 
leaves a residual unfixed $H$ gauge invariance.
The defect manifold $N$ of the Abelian gauge is formed by those points 
$x$ of space time where $\phi(A^{\bar g})(x)$ is invariant under 
the ajoint action of a subgroup $K$ of $G$ properly containing $H$.
$\bar g$ is then singular at a Dirac manifold $D$ bounded by $N$. 
$\bar A=A^{\bar g}$ is also singular at $D$.
A center gauge works in similar fashion, but it requires two linearly independent 
Higgs fields $\phi(A)$, $\phi'(A)$ satisfying (1.1) rather than a single one. 
The gauge transformation $g$ which carries $A$ into the center gauge 
is the one which renders $\phi(A^g)$, $\phi'(A^g)$ $\goth h$, $\goth v$  
valued, respectively, where $\goth v$ is a proper subspace of $\goth g$ 
not contained $\goth h$. $g$ is defined up to right 
multiplication by an arbitrary element $z$ of $Z$, $g\rightarrow gz$.
So, the Abelian gauge fixing leaves a residual unfixed $Z$ gauge invariance.
$g$ becomes singular at those points in space time where the residual $Z$
gauge symmetry is enhanced to a larger subgroup $W$ of $G$ containing $Z$. 
In either types of gauges, the the gauge fixed gauge field 
$A^g$ is singular at the defects.


There is an interesting relation between monopole and instanton charge, 
which can be seen as follows. We assume again that the open cover $\{O_i\}$ 
is good. As $f_{\scri C}\in\Omega^2_c(M)$,  one has a \v Cech--de Rham 
descent
$$
f_{\scri C}|_{O_i}=dv_i,\quad\delta v_{ij}=dw_{ij}, 
\quad \delta w_{ijk}=m_{ijk}, \quad \delta m_{ijkl}=0 
,\eqno(7.11)
$$
where $\{v_i\}$ is an $\Omega^1$ \v Cech $0$--cochain,
$\{w_{ij}\}$ is an $\Omega^0$ \v Cech $1$--cochain and
$\{m_{ijk}\}$ is a $2\pi\Bbb Z$ $2$cochain.
If $\Omega\in Z^s_4(M)$ is an $\{O_i\}$--small finite singular 4--cycle, 
then one has the \v Cech--singular descent
$$
\Omega=\sum_iU_i, \quad bU_i=\sum_jV_{ji}, \quad bV_{ij}=\sum_kW_{kij},
\quad bW_{ijk}=\sum_lX_{lijk},
\eqno(7.12)
$$
where $\{U_i\}$ is a $C^s_4$ \v Cech $0$--chain, 
$\{V_{ij}\}$ is a $C^s_3$ \v Cech $1$--chain,
$\{W_{ijk}\}$ is a $C^s_2$ \v Cech $2$--chain and 
$\{X_{ijkl}\}$ is a $C^s_1$ \v Cech $3$--chain.
Using the descent (7.11) for one of the factors of 
$f_{\scri C}{}^2|_{O_i}$ and using Stokes' theorem repeatedly, 
one finds that
$$
i_{\scri C}(\Omega)=\hbox{$1\over 2$}m_{\scri C}(\Theta_\Omega)
,\eqno(7.13)
$$
where $\Theta_\Omega\in C^s_2(M)$ is given by
$$
\Theta_\Omega=\hbox{$1\over 6$}\sum_{ijk}(m_{ijk}/2\pi)W_{ijk}
.\eqno(7.14)
$$
Using (7.11), (7.12), it is easy to show that $\Theta_\Omega\in Z^s_2(M)$ 
is actually an $\{O_i\}$--small finite singular 2--cycle. One can further 
show that the (co)homological ambiguities of the descents (7.11),
(7.12) affect $\Theta_\Omega\in Z^s_2(M)$ by a singular $2$--boundary.
Finally, it is apparent that $\Theta_\Omega$ is $\Bbb Z$ linear in $\Omega$
up to singular $2$--boundaries. Using barycentric subdivision, 
the above calculation can be generalized to non necessarily 
$\{O_i\}$--small finite singular 4--cycles $\Omega\in Z^s_4(M)$.
$[\Theta_\Omega]=\mu\cap[\Omega]$ where $\mu\in H^2(M,\Bbb Z)$ 
is the $\Bbb Z$ degree 2 cohomology class defiend by the
$\Bbb Z$ \v Cech $2$--cocycle $\{m_{ijk}/2\pi\}$ and $\cap $
denotes the cap product. $\mu$ is determined by $f_{\scri C}$
only up to torsion and so is $[\Theta_\Omega]$. 
(7.13), (7.14) furnishes a cohomological
justification of the relation of monopole and instanton charge found in 
ref. \ref{36}, though the counterpart of twist in the present formulation 
is not clear.

An $\SO(3)$ vector bundle $F$ on $M$ has an $\SO(2)$ 
reduction $F=G\oplus\ul{\Bbb R}$, where $G$ is an $\SO(2)$ vector bundle on $M$ and 
$\ul{\Bbb R}=\Bbb R\times M$, whenever there is a nowhere vanishing section 
$\ul v\in\Omega^0(M,F)$. $G$ is the subbundle of $F$ of local sections
$\ul s$ such that $\ul s\cdot\ul v=0$. A connection $\ul C\in\Conn(M,F)$ of an 
$\SO(3)$ vector bundle $F$ is $\SO(2)$ reducible if there is a non trivial section 
$\ul w\in\Omega^0(M,F)$ such that $\ul D_{\ul C}\ul w=0$. 
If $\ul w$ is nowhere vanishing, then $\ul w$ induces an $\SO(2)$ reduction of 
$F$ with $G$ invariant under parralel transport by $\ul C$

In general, a given Cho structure 
$\scri C$ is not fine. By replacing $\scri C$ with an equivalent Cho structure, 
one may try to enforce fineness. However, the analysis of specific examples shows 
that, in general, it is not possible to do that globally on $M$, but only upon 
restricting to an open subset $M\setminus D$ of $M$, where $D$ is a submanifold 
of $M$ of non zero codimension. Indeed, $E_{\scri C}$ is singular along 
$D$ and the $\ul A_{\scri C}$ and $\ul n_{\scri C}$ are singular along 
$N=\partial D$. $N$ may be called the {\it defect manifold} of $\scri C$

In physical applications, $P$ is a principal $\SU(2)$ bundle defined by the $\SU(2)$ 
valued monodromy matrix functions of the gauge fields. $M$ is some $4$--dimensional 
space--time manifold, such as $\Bbb R^4$, $\Bbb S^4$, $\Bbb S^1\times\Bbb R^3$, 
$\Bbb T^4$, $D$ is a set of Dirac sheets in $M$ and $N$ is a set of monopole world 
lines forming a knot in $M$ (See refs. \ref{32,33} for a detailed physical discussion 
of this point.)

\titlecps{Examples}

We now illustrate the above analysis with a few examples.

\exa {3.5} {\it A small monopole loop in a single instanton background}

The authors of ref. \ref{34} found a solution of the differential maximal 
Abelian gauge in a single instanton background. Its defect manifold is a 
loop. When the radius of the loop is much smaller than the instanton size, 
an analytic expression is available. In this case, $M$ is the $4$--space  
$\Bbb R^4$ and $P$ is the trivial principal bundle $\Bbb R^4\times \SU(2)$. 
$E=\Bbb R^4\times\Bbb E_3$. 

We define coordinates $u,~v\in[0,+\infty[$, $\varphi,~\psi\in[0,2\pi[$ 
in $\Bbb R^4$ by
$$
x^1+\sqrt{-1}x^2=u\exp\big(\sqrt{-1}\varphi\big), \quad
x^3+\sqrt{-1}x^4=v\exp\big(\sqrt{-1}\psi\big).
\eqno(3.7)
$$
Note that $\phi$, $\psi$ are ill defined at the planes $u=0$, $v=0$,
respectively. We set 
$$
\eqalignno{
\vphantom{1\over 2}
O_1&=\{x|x\in\Bbb R^4,~u>R_1\hbox{~or~}v>R_1\}, &(3.8)\cr
\vphantom{1\over 2}
O_2&=\{x|x\in\Bbb R^4,~u<R_2\hbox{~and~}v<R_2\}, &\cr
}
$$
where $R_2>R_1>0$. $\{O_1,O_2\}$ is an open covering of $\Bbb R^4$.

Let $\alpha:\Bbb R^4\rightarrow[0,2\pi[$, $\beta:\Bbb R^4\rightarrow[0,\pi]$,
$\gamma:\Bbb R^4\rightarrow[0,2\pi[$ be given by
$$
\eqalignno{\vphantom{1\over 2}
\alpha&=\varphi-\psi \quad\hbox{mod $2\pi$}, &(3.9a)\cr
\vphantom{1\over 2}
\beta &=\tan^{-1}\Big({2uv\over u^2-v^2-R_0{}^2}\Big)
\quad\hbox{mod $\pi$}, &(3.9b)\cr
\vphantom{1\over 2}
\gamma&= \varphi+\psi \quad\hbox{mod $2\pi$},&(3.9c)\cr
}
$$
where $R_2,~R_1>R_0>0$. $\alpha$, $\gamma$ are ill defined at the planes 
$u=0$, $v=0$, $\beta$ is ill defined at the hypersurfaces $u^2-v^2-R_0{}^2=0$.
Let $s_1\in\Omega^1(O_1)$, $s_2\in\Omega^1(O_2)$, $\chi_{12}\in\Omega^0(O_{12})$ 
be such that 
$$
s_2-s_1=-d\chi_{12}.
\eqno(3.10)
$$
We set
$$
\eqalignno{\vphantom{1\over 2}
\ul n_1&=\big[\sin\beta(\cos\alpha\,\ul e_1+\sin\alpha\,\ul e_2)
+\cos\beta\,\ul e_3\big]\big|_{O_1}, &(3.11)\cr
\vphantom{1\over 2}
\ul n_2&=\ul e_3, &\cr
\vphantom{1\over 2}
\ul A_1&=\big[-d\alpha\,\ul e_3
-d\beta\big(-\sin\alpha\,\ul e_1+\cos\alpha\,\ul e_2\big)&(3.12)\cr
\vphantom{1\over 2}
&\hphantom{=}~
+(\cos\beta d\alpha-s_1)\big(\sin\beta(\cos\alpha\,\ul e_1+\sin\alpha\,\ul e_2)
+\cos\beta\,\ul e_3)\big]\big|_{O_1}, &\cr
\vphantom{1\over 2}
\ul A_2&=-s_2\ul e_3, &\cr
}
$$
$$
\eqalignno{
\vphantom{1\over 2}
\ul T_{12}&=\exp(-\alpha\star\ul e_3)\exp(-\beta \star\ul e_2)
\exp(-(\gamma+\chi_{12})\star\ul e_3)\big|_{O_{12}}, &(3.13)\cr
\vphantom{1\over 2}
\ul T_{21}&=\exp((\gamma+\chi_{12})\star\ul e_3)\exp(\beta \star\ul e_2)
\exp(\alpha\star\ul e_3)\big|_{O_{12}}, &\cr
\vphantom{1\over 2}
\ul T_{11}&=\ul T_{22}=\ul 1. &\cr
}
$$
If we neglect the restriction to $O_1$, $\ul n_1$, $\ul A_1$ are regular everywhere 
exept on the circle 
$$
N=\{x|x\in\Bbb R^4,~u=R_0,~v=0\}.
\eqno(3.14)
$$
Likewise, if we neglect the restriction to $O_{12}$, $\ul T_{12}$, $\ul T_{21}$
are are regular everywhere exept on the disk 
$$
D=\{x|x\in\Bbb R^4,~u\leq R_0,~v=0\}
\eqno(3.15)
$$
bounded by $N$. Note that $N\cap O_1=\emptyset$, $N\subseteq O_2$ and that
$D\cap O_{12}=\emptyset$.
It is then straightforward though lengthy  
to check that (3.11)--(3.13) define a Cho structure $\scri C$ of 
$E$ subordinated to $\{O_1,O_2\}$ and that $N$ is its defect manifold.

\exa {3.6} {\it A Higgs field on a 4--torus}

In this case, $M$ is the $4$--torus $\Bbb T^4$ and $P$ is the trivial 
principal bundle $\Bbb T^4\times \SU(2)$. $E=\Bbb T^4\times \Bbb E_3$. 

We view $\Bbb T^4$, as the quotient $\Bbb R^4$ by a lattice 
$\Lambda\subseteq \Bbb R^4$, $\Bbb T^4=\Bbb R^4/\Lambda$. 
Let $\{O_i\}$ an open covering of $\Bbb T^4$
such that, for each $i$, $O_i$ is a simply connected non empty open subset of 
$\Bbb T^4$. As is well known, for each $i$, there is a local coordinate 
$x_i$ of $\Bbb T^4$ defined on $O_i$ such that 
$$
\theta=x_i(\theta)+\Lambda,
\eqno(3.16)
$$
for $\theta\in O_i$. Further, when $O_{ij}\not=\emptyset$, there is 
$\xi_{ij}\in\Lambda$ such that
$$
x_i=x_j+\xi_{ij}\quad \hbox{on $O_{ij}$}
\eqno(3.17)
$$
The collection $\{\xi_{ij}\}$ is a $\Lambda$ valued \v Cech 
$1$--cocycle.

Let $c:\Bbb T^4\rightarrow\Bbb R^{4\vee}$ be a smooth function
and $\ul k_0,~\ul e_0\in \Bbb E_3$ be unit vectors.
Let $s_i$, $\exp\big(\sqrt{-1}\chi_{ij}\big)$ be
the local representatives of a connection $s$ and the transition function 
of some $\Bbb T$ principal bundle on $M_0$, respectively, so that 
$$
s_j-s_i=-d\chi_{ij}
.\eqno(3.18)
$$
We set 
$$
\eqalignno{\vphantom{1\over 2}
\ul n_i&=\exp\big(-\langle c, x_i\rangle\star \ul k_0\big)\ul e_0,
&(3.19)\cr
\vphantom{1\over 2}
\ul A_i&=-\exp\big(-\langle c, x_i\rangle\star \ul k_0\big)
\big[d\langle c, x_i\rangle\ul e_0\times(\ul k_0\times\ul e_0)+s_i\ul e_0\big],
&(3.20)\cr
\vphantom{1\over 2}
\ul T_{ij}&=\exp\big(-\langle c, x_i\rangle\star \ul k_0\big)
\exp\big(-\chi_{ij}\star \ul e_0\big)
\exp\big(\langle c, x_j\rangle\star \ul k_0\big)|_{O_{ij}}.&(3.21)\cr
}
$$
It is straightforward to check that (3.19)--(3.21) define a Cho structure 
${\scri C}$ of $E$ subordinated to $\{O_i\}$.
Its defect manifold $N$ is empty, $N=\emptyset$.

\vskip .3cm 
$\underline{\hbox{\cps Discussion}}$
\vskip .3cm

Cho structures provide a local formulation of Cho's original gauge theoretic
formalism \ref{24,25}.

We recall that a connection $\ul C\in\Conn(M.F)$ of an $\SO(3)$ vector bundle 
$F$ is $\Bbb T$ reducible if and only if there is a non trivial section 
$\ul s\in\Omega^0(M,F)$ such that $\ul D_{\ul C}\ul s=0$ \ref{31}. 

From (3.1), (3.2), it follows that a Cho structure $\scri C$ of $E$ is a family 
of local $\Bbb T$ reducible connections of $E$ subordinated to the associated 
covering. 

In general, the local data of a Cho structure cannot be assembled in a global 
structure basically because the $\ul T_{ij}$ do not necessarily satisfy 
the 1--cocycle relation 
$$
\ul T_{ij}\ul T_{jk}=\ul T_{ik}  \quad\hbox{(false in general)}.
\eqno(3.7)
$$
However, if they do, the $\SO(E)$ 1--cocycle $\{\ul T_{ij}\}$ defines 
a new $\SO(3)$ vector bundle $F$ by pasting. Further, 
the $\Omega^0(E)$ 0--cochain $\{\ul n_i\}$ defines a non trivial section 
$\ul s\in\Omega^0(M,F)$. If, moreover,
$$
\ul A_i=\ul T_{ij}(\ul A_j-\ul\zeta_{ij}) \quad\hbox{(false in general)},
\eqno(3.8)
$$
where $\star\,\ul \zeta_{ij}=-\ul T_{ij}{}^{-1}d\ul T_{ij}$, then 
the $\Conn(E)$ 0--cochain $\{\ul A_i\}$ defines a connection $\ul C\in\Conn(M,F)$.
Further, 
$$
\ul D_{\ul C}\ul s=0
\eqno(3.9)
$$
Hence, when (3.7), (3.8) hold, a Cho structure defines an $\SO(3)$ vector bundle $F$ 
with a $\Bbb T$ reducible connection $A$. However, in general, (3.7), (3.8) fail 
to hold. 

By replacing the given Cho structure $\scri C$ with an equivalent one
${\scri C}'$, one may try to enforce (3.7), (3.8). However, it may not be 
possible to do that globally on $M$. In general, there appears a defect submanifold 
$N$ of $M$ where either $\ul s$ or $\ul C$ become singular; (3.7), (3.8) then hold 
only on $M\setminus N$.

In physical applications, $P$ is a principal $\SU(2)$ bundle defined by the $\SU(2)$ 
valued monodromy matrix functions of the gauge fields.
$M$ is some $4$--dimensional space--time manifold, 
such as $\Bbb R^4$, $\Bbb S^4$, $\Bbb S^1\times\Bbb R^3$, $\Bbb T^4$,
$N$ is a set of monopole world lines forming a knot in $M$ \ref{32,33}.

We recall that a connection $\ul C\in\Conn(M,F)$ of an $\SO(3)$ vector bundle 
$F$ is $\Bbb T$ reducible if and only if there is a non trivial section 
$\ul s\in\Omega^0(M,F)$ such that $\ul D_{\ul C}\ul s=0$ \ref{31}. 

From (3.1), (3.2), it follows that a Cho structure $\scri C$ of $E$ is a family 
of local $\Bbb T$ reducible connections of $E$ subordinated to the associated 
covering. 

In general, the local data of a Cho structure cannot be assembled in a global 
structure basically because the $\ul T_{ij}$ do not necessarily satisfy 
the 1--cocycle relation

However, if they do, the $\SO(E)$ 1--cocycle $\{\ul T_{ij}\}$ defines 
a new $\SO(3)$ vector bundle $F$ by pasting. Further, 
the $\Omega^0(E)$ 0--cochain $\{\ul n_i\}$ defines a non trivial section 
$\ul s\in\Omega^0(M,F)$. If, moreover,

then 
the $\Conn(E)$ 0--cochain $\{\ul A_i\}$ defines a connection $\ul C\in\Conn(M,F)$.
Further, 
$$
\ul D_{\ul C}\ul s=0
\eqno(3.9)
$$
Hence, when (3.7), (3.8) hold, a Cho structure defines an $\SO(3)$ vector bundle $F$ 
with a $\Bbb T$ reducible connection $A$. However, in general, (3.7), (3.8) fail 
to hold. 

\titlecps{Discussion}

By replacing the given Cho structure $\scri C$ with an equivalent one
${\scri C}'$, one may try to enforce (3.7), (3.8). However, it may not be 
possible to do that globally on $M$. In general, there appears a defect submanifold 
$N$ of $M$ where either $\ul s$ or $\ul C$ become singular; (3.7), (3.8) then hold 
only on $M\setminus N$.

In physical applications, $P$ is a principal $\SU(2)$ bundle defined by the $\SU(2)$ 
valued monodromy matrix functions of the gauge fields.
$M$ is some $4$--dimensional space--time manifold, 
such as $\Bbb R^4$, $\Bbb S^4$, $\Bbb S^1\times\Bbb R^3$, $\Bbb T^4$,
$N$ is a set of monopole world lines forming a knot in $M$ \ref{32,33}.

The standard framework in which Cho structures are defined 
can be described as follows \ref{31,32}.
Consider a $4$--dimensional manifold $M_0$ and 
a closed $1$--dimensional submanifold $N$ of $M$ and set 
$$
M=M_0\setminus N
,\eqno(3.7)
$$
Consider further a principal $\SU(2)$ bundle $P_0$ on $M_0$ and set
$$
P=P_0|_M
.\eqno(3.8)
$$
The vector bundle $E$ is then defined according (2.2).

Let ${\scri C}=(\{\ul n_i\},\{\ul A_i\},\{\ul T_{ij}\})$ be a Cho
structure of $E$ subordinated to the cover $\{O_i\}$ of $M$. 
In general, if $\partial O_i$ intersects $N$, $\ul n_i$,
$\ul A_i$ and $\ul T_{ij}$ become singular on approaching $N$. 
$N$ thus is their defect manifold.

It is important to realize that 
the constructions worked out in this paper are fully general and do not require 
that $M$ and $P$ are of the form (3.7), (3.8) or that $\dim M=4$.

\def\defn#1{\vskip.2cm\par\noindent$\underline{\hbox{\it Definition {#1}}}$
\par\noindent}
\def\prop#1{\vskip.2cm\par\noindent$\underline{\hbox{\it Proposition {#1}}}$
\par\noindent}
\def\defprop#1{\vskip.2cm\par\noindent$\underline{\hbox{\it Definition and 
Proposition {#1}}}$
\par\noindent}
\def\theo#1{\vskip.2cm\par\noindent$\underline{\hbox{\it Theorem {#1}}}$
\par\noindent}

\def\rema#1{\vskip.2cm\par\noindent$\underline{\hbox{\it Remark {#1}}\vphantom{f}}$
\par\noindent}
\def\proof{\vskip.2cm\par\noindent$\underline{\hbox{\it Proof}}~~$}
\def\proclaim #1. #2\par{\medbreak{\it #1.\enspace}{#2}\par\medbreak}

\vfill\eject

In this case, $M_0$ is the $4$--manifold $\Bbb S^1\times\Bbb R^3$ and $P_0$ 
is the trivial principal bundle $(\Bbb S^1\times\Bbb R^3)\times\SU(2)$. 
The defect manifold $N$ is the loop $\Bbb S^1\times\{0\}$.
Thus, $M=\Bbb S^1\times(\Bbb R^3\setminus\{0\}$, 
$P=(\Bbb S^1\times(\Bbb R^3\setminus\{0\})\times\SU(2)$ and 
$E=(\Bbb S^1\times(\Bbb R^3\setminus\{0\})\times\Bbb E_3$. 

We describe $\Bbb S^1\times\Bbb R^3$ by an angular coordinate $\tau\in[0,2\pi]$
for the factor $\Bbb S^1$ and standard spherical coordinates 
$r\in[0,+\infty[$, $\vartheta\in[0,\pi[$, $\varphi\in[0,2\pi]$
for the factor $\Bbb R^3$. Rexall that  the latter are ill defined at the line 
$x^1=x^2=0$. We have
$$
N=\{x|x\in\Bbb S^1\times\Bbb R^3,~r=0\}.
\eqno(3.7)
$$
We set next 
$$
O_1=\{x|x\in\Bbb S^1\times\Bbb R^3,~r\not=0\}. 
\eqno(3.8)
$$
Note that $O_1=M$. $\{O_1\}$ is an open covering of $M$, trivially.

Let $m,k\in\Bbb Z$. We set
$$
\eqalignno{\vphantom{1\over 2}
\ul n_1&=\big[\sin\vartheta(\cos(m\varphi-k\tau)\,\ul e_1
+\sin(m\varphi-k\tau)\,\ul e_2)
+\cos\vartheta\,\ul e_3\big]\big|_{O_1}, &(3.14)\cr
\vphantom{1\over 2}
\ul A_1&=\big[-d(m\varphi-k\tau)\,\ul e_3
-d\vartheta\big(-\sin(m\varphi-k\tau)\,\ul e_1+\cos(m\varphi-k\tau)\,\ul e_2\big)
&(3.15)\cr
\vphantom{1\over 2}
&\hphantom{=}~
+\cos\vartheta\,d(m\varphi-k\tau)\big(\sin\vartheta(\cos(m\varphi-k\tau)\ul e_1
+\sin(m\varphi-k\tau)\ul e_2)+\cos\vartheta\,\ul e_3)\big]\big|_{O_1}, &\cr
\vphantom{1\over 2}
\ul T_{11}&=\ul 1. &(3.16)\cr
}
$$
It is straightforward to verify that these objects are singular on the circle $N$
and nowhere else, as they should. It is also simple to check that
(3.14)--(3.16) define a Cho structure $\scri C$ of $E$ subordinated to $\{O_1\}$.

The only non vanishing component of the Deligne $3$--cocycle associated with 
the Cho structure $\scri C$ (3.14)--(3.16) is
$$
\eta_1=\big[-\sin\vartheta\,d(m\varphi-k\tau)d\vartheta\big]\big|_{O_1}
\eqno(4.26)
$$

\bye